\documentstyle[psfig,12pt]{article}
\begin{document}
\title{A Theory of Non-Gaussian Option Pricing}
\author{Lisa Borland \\ Evnine-Vaughan Associates, Inc.\\456 Montgomery Street, Suite 800\\ San Francisco, CA 94104-1241,  USA}
\maketitle
\begin{abstract}
Option pricing formulas are derived from a non-Gaussian model of stock returns.
Fluctuations are assumed to  evolve 
according to a nonlinear Fokker-Planck equation which maximizes the Tsallis nonextensive
entropy of index $q$.  A generalized form of
the Black-Scholes differential equation is found, and 
we derive  a  martingale measure which leads to closed form solutions for European 
call options.  The standard Black-Scholes pricing 
equations are recovered  as a special case ($q = 1$). The distribution of stock returns is
well-modelled with $q$ circa $1.5$. Using that value of $q$ in the option pricing model  we 
reproduce the volatility smile. The partial
derivatives (or Greeks) of the model
are also  calculated. Empirical results are demonstrated for options on Japanese Yen futures.
Using just one value of $\sigma$ across strikes we closely reproduce 
market prices,  for  expiration times ranging from  weeks
 to several months.  
\end{abstract}

{ \footnotesize{\bf Acknowledgements} Many insightful discussions with Roberto Osorio and Jeremy 
Evnine are gratefully acknowledged.}

\newpage
\section{Introduction}
It is well known that the distributions of empirical returns do not follow
the lognormal distribution upon which many celebrated results of finance are
 based. For example, Black and Scholes \cite{blackscholes} and  Merton \cite{merton}
 were able to derive the
  prices of options and other derivatives of the
 underlying stock based on such a model. While of great importance and widely used,
 such theoretical option prices do not quite
match the observed ones. In particular, the Black-Scholes model underestimates the prices
of away-from-the-money options. This means that the  implied volatilities of
options of various strike prices form a convex function, rather than the expected flat line.
This is known as the ``volatility smile''.

Indeed, there have been several modifications to the standard models in an 
attempt to correct for these discrepencies.
One approach is to introduce a stochastic model for the volatility of the stock price,
as was done by Hull and White \cite{Hull&White}, or via a
generalized autoregressive conditional heteroskedasticity (GARCH) model of volatility,
A review and references can be found in \cite{Hull}.  
Another class of models include a  Poisson jump diffusion term \cite{Merton76} which can describe extreme price movements.  
The DVF (Deterministic Volatility Function) approach \cite{Dupire}, as well as  combinations of some of these different approaches \cite {jesper}, have also been studied. 
A quite different line of thought  is offered in \cite{bouchaud1,bouchaud}, where it is argued that  
heavy non-Gaussian 
tails and finite hedging time make it necessary to go beyond the notion of  risk-free option prices.
They obtain non-unique prices, associated with a given level of risk.
More recently, other techniques along the lines of  \cite{hyperbolic} lead to  option
prices  based on an underlying hyperbolic distribution.

In many cases,  these approaches are often either
very complicated or rather ad-hoc.  To our knowledge, none result in  managable closed form
solutions, which is a useful result  of the Black and Scholes approach. In this paper
we do however succeed in obtaining closed form solutions for European options.
Our  approach is based on a new class of stochastic processes which allow for statistical
 feedback as a model of the underlying
 stock returns. 
 We can show  that the distributions of returns implied by 
these processes closely match those found empirically. In particular they
 capture features such as the fat tails and peaked middles which are not at all
captured by the standard class of lognormal distributions.

 Our stochastic model 
derives from a class
of processes \cite{pdependent} which  have 
been  recently developed within the framework of statistical 
physics, namely within the very active  field of Tsallis nonextensive thermostatistics 
\cite{Tsallis}. Many interesting applications of this new statistical
 paradigm have been found 
in recent years,  mainly  related to the sciences, although there
are some results showing that the power-law distributions characteristic
of the Tsallis framework are  good models for the distributions of certain 
financial quantities \cite{examples,Michael&Johnson,osorio}. However, to our knowledge   
the current work, a short version of which is given in \cite{shortpaper},  contains the 
first application  of the
 associated  stochastic  processes to finance.  

Basically, these stochastic processes can be interpreted as if the driving noise follows
a generalized Wiener process governed by a fat-tailed Tsallis distribution of index $q > 1$.
For $q=1$ the Tsallis distribution coincides with a Gaussian and the standard 
stock-price model is recovered. However, for $q>1$ these distributions exhibit
fat tails and
  appear to be  good models of real data, as shown in Figure 1.
There, the empirical distribution of 
the log daily price returns (ignoring dividends and non-trading days) to 
the  de-meaned S\&P 500 is plotted.  Returns were normalized by the  sample standard
 deviation of the
series which is   19.86 \% annualized, and then binned.
For comparison,   the distribution  obtained from  a Tsallis 
distribution of index $q=1.43$ is also plotted \cite{osorio}. 
It seems clear that the Tsallis distribution  provides  a
 much better fit to the empirical distribution than the lognormal, which is also shown. 
Another example is shown in Figure 2, where  the distribution of high frequency log returns for 10 Nasdaq  high-volume  
stocks is plotted \cite{osorio}. The timescale is 1 minute. Again, returns are normalized by the sample standard  deviation. A Tsallis distribution of index $q = 1.43$ provides a very good fit to the empirical data.
Another  example  of  such  a match 
between Tsallis distributions ($q=1.6$) and those of financial returns over different timescales can  be found in \cite{Michael&Johnson}.

Motivated by the  good fit between the proposed model class and empirical data,
we use these stochastic processes to represent movements of the returns of the 
 underlying stock. We then   derive generalized option
 pricing formulas so 
as to be able to obtain fair values of derivatives of the underlying.  
Using these formulas we get a good match with empirically observed option prices.
In particular, we show in this paper (see Figures 16 and 17) that a $q=1.4$ model with 
one value of $\sigma$ across 
strikes reproduces market prices for options on Japanese Yen futures
with expiry dates ranging from 17 to 147 days.

\section{ The Model of Returns}
The standard model for stock price movement is that
\begin{equation}
S(\tau + t) = S(\tau) e^{Y(t)}
\end{equation}
where $Y$ follows the stochastic process
\begin{equation}
\label{eq:lanstand}
dY = \mu dt + \sigma d\omega
\end{equation}
The drift $\mu$ is the mean rate of return and $\sigma^2$ is the variance of the stock logarithmic return. 
 The driving noise 
$\omega$ is a Brownian motion defined with respect to a probability measure $F$. It represents a  Wiener process and 
has the property 
\begin{equation} 
\label{eq:noiseproperty}
E^F[d\omega(t)d\omega(t')] = dtdt'\delta(t-t')
\end{equation}
where the notation  $E^F[]$ means the expectation value with respect to the measure $F$.
Note that  the conditional probability distribution of the variable $\omega$ 
satisfies the Fokker-Planck equation
\begin{equation}
 \frac{\partial P(\omega ,t \mid \omega ', t')}{\partial t} =  \frac{1}{2} \frac{\partial ^2}{\partial \omega^2} P(\omega ,t  \mid \omega ', t'))
\end{equation}
and is distributed according to
\begin{equation}
\label{eq:pomega}
P(\omega(t), t \mid \omega (t'), t') = \frac{1}{\sqrt{2 \pi (t-t')}} \exp (- \frac{(\omega(t) - \omega(t')) ^2}{2(t-t')})
\end{equation}
In addition one chooses   $t' = 0$ and  $\omega(0) = 0$  so that this  defines a Wiener process, which is distributed according to a zero-mean Gaussian.

It is well-known that this model gives a normal distribution with drift $\mu t$ and variance $\sigma^2 t$ for the variable
 $Y$.
This can for example be seen by rewriting Eq (\ref{eq:lanstand}) as 
\begin{equation}
 d(\frac{ Y - \mu t }{\sigma} ) = d \omega 
\end{equation}
which indicates that we can substitute 
\begin{equation}
\label{eq:omegatos}
\omega = (Y - \mu t)/\sigma
\end{equation}
 into Eq (\ref{eq:pomega}). We obtain the well-known lognormal distribution
for the stock returns over timescale $T$, after inserting $Y = \ln S(\tau + t)/S(\tau)$ :
\begin{equation}
\label{eq:pgaussian}
P(\ln S(T + \tau) \mid \ln S(\tau)) = N  \exp ( - \frac{ ( \ln \frac{S(\tau + T)}{S(\tau)} - \mu T ) ^2} {2 \sigma^2 T })
\end{equation}

Based on this stock-price model, Black and Scholes were able to establish 
a pricing model  to obtain the fair value of options on the underlying stock $S$.
However, their model predicts a lognormal distribution, whereas empirical distributions of stock returns are better fitted with power-law distributions \cite{bouchaud,cubiclaw,Michael&Johnson,osorio}. Here we focus mainly on the empirical evidence of \cite{Michael&Johnson,osorio} where it is shown that the distributions which naturally arise within the  framework of the generalized thermostatistics of Tsallis \cite{Tsallis} provide very good fits to empirical distributions of returns on different timescales.

  In contrast to other models where the standard Black-Scholes price model is extended to account for non-normal noise, for example jump diffusion models \cite{Merton76} and Levy noise \cite{hyperbolic}, we introduce here a new model of stock return fluctuations, which derives directly from stochastic processes recently introduced within the Tsallis framework \cite{pdependent}.
  In this setting, we assume that 
the log returns $Y(t) = \ln S (\tau+t)/\ln S(\tau)$ follow the process
\begin{equation}
\label{eq:lancondY}
d Y = \mu dt + \sigma d \Omega
\end{equation}
across timescales $t$,
where  we shall now model the driving noise $\Omega$  as being drawn from a
 non-Gaussian 
distribution. To do this, we assume that $\Omega$
 follows the statistical feedback process \cite{pdependent}
\begin{equation}
\label{eq:nomega}
d \Omega =  P(\Omega)^{\frac{1-q }{2}} d \omega
\end{equation}
Here $\omega$ is a zero-mean Gaussian noise process as defined above. For $q = 1$,
$\Omega$ reduces to $\omega$ and the standard model is recovered.
The probabilty distribution of the variable $\Omega$ evolves according to the 
nonlinear Fokker-Planck equation \cite{pdependent}
\begin{equation}
\label{eq:nlfpcond}
\frac{ \partial}{\partial t} P(\Omega, t \mid \Omega ', t') = \frac{1}{2} \frac{ \partial }{\partial \Omega ^2} P^{2-q}(\Omega, t \mid \Omega ', t')
 \end{equation}
It can be verified that the  conditional probability  $P$ that solves this system is given by so-called Tsallis distributions (or q-Gaussians $P_q$)
\begin{equation}
\label{eq:ptsalliscond}
P_q(\Omega, t \mid \Omega', t') = \frac{1}{Z(t)} \left(1 - \beta(t) (1-q)(\Omega - \Omega') ^2 \right)^{\frac{1}{1-q}}
\end{equation}
with 
\begin{equation}
\label{eq:betacond}
\beta(t) = c^{\frac{1-q}{3-q}}((2-q) (3-q) (t-t'))^{-2/(3-q)}
\end{equation} 
and
\begin{equation}
\label{eq:Z}
Z(t) =  ( (2-q) (3-q) c (t-t') )^{\frac{1}{3-q}} 
\end{equation} 
By choosing $t'=0$ and  $\Omega(0) = 0$, we obtain a generalized Wiener 
process, distributed according to a zero-mean Tsallis distribution
\begin{equation}
\label{eq:ptsalliscondzeromean}
P_q(\Omega, t \mid 0, 0) = \frac{1}{Z(t)} \left(1 - \beta(t) (1-q)(\Omega ) ^2 \right)^{\frac{1}{1-q}}
\end{equation}
The index $q$ is known as the entropic index of the generalized Tsallis
 entropy.  The $q$-dependent constant $c$ is given by
\begin{eqnarray}
c & = & \beta Z^2 \\
Z & = &\int_{- \infty}^{\infty} ( 1 - (1-q)\beta \Omega^2)^{\frac{1}{1-q}} d \Omega
\end{eqnarray}
for any $\beta$. 
In the limit $q \rightarrow 1$ the standard theory is recovered, 
and $P_q$ becomes a Gaussian.
In that case, the standard Gaussian driving noise of Eq (\ref{eq:lanstand}) 
is also recovered.
 For $q < 1$  these distributions
 exhibit a so-called cutoff resulting in regimes where $P_q=0$. In the current
 paper, we will therefore only consider values of $q > 1 $, for
 which the distributions exhibit fat tails. There is also a natural limit  
 at $q = 3$ after which value the distributions are no longer 
normalizable.  Another important point which constrains the 
realistic range of $q$-values is the fact that the variance of the Tsallis 
distributions is given by \cite{Levy}
\begin{equation}
\label{eq:qvariance}
 E[\Omega^2(t)] =  \frac{1}{( 5 - 3q) \beta(t)}
\end{equation}
Clearly, this expression diverges for $q \ge 5/3$. Since we are only interested in processes with finite variance, we assume $1 < q < 5/3$, which covers the
values of empirical interest.

Our  model exhibits a 
 statistical feedback into the system, from the macroscopic level characterised
 by $P$, to the microscopic level characterised by the dynamics of $\Omega$, and 
thereby ultimately by the returns $Y$.
This scenario is simply a phenomenological 
description of the underlying dynamics. For example, in the case of stock 
prices, we can imagine that the statistical feedback is really due to the 
interactions of many individual traders whose actions all will contribute to
shocks to  the stock price which  keep it in equilibrium. Their collective behaviour
can be summarized by the statistical dependency in the noise term of
the stochastic model for $Y$. This yields a nonhomogenous
 reaction to the returns: depending on the value of $q$, rare events (i.e.
 extreme returns) will be accompanied by large reactions.
 On the  other hand, if the returns take  on less extreme values,   then the 
size of the noise  is more moderate.

The time-dependent solutions presented here can be seen 
as a special case of 
those presented in \cite{Tsallis&Bukman} where the $\delta$-function as initial condition was not explicitly  discussed.
It is not difficult to verify that the particular form of $P_q$ which we 
introduce  here has the 
property that $P_q$ becomes sharply peaked as $t$ approaches zero. In other words, it
approaches a $\delta$-function  as $t\rightarrow 0$, which corresponds to the 
fact that the returns are known with  certainty to be  zero over intervals   $t = 0$.

Let us look at what 
effect the driving noise $\Omega$ has on the log returns $Y(t) $. We can write
\begin{equation}
\label{eq:dlnsom}
d ( \frac{Y - \mu t}{\sigma} ) = d \Omega 
\end{equation}
or equivalently 
\begin{equation}
\label{eq:omtolns}
\Omega(t)  = (\ln \frac{S(\tau +t)}{S(\tau)} - \mu t )/\sigma
\end{equation}
It then follows  from  Eq (\ref{eq:ptsalliscond}) that  the distribution of returns $\ln S(\tau + t)/S(\tau)$ obeys 
\begin{equation}
\label{eq:ptsallisnonzero}
P_q(\ln S (\tau + t) \mid \ln S(\tau)) = \frac{1}{Z(t)} \left(1 - \tilde{\beta}(t) (1-q)(\ln \frac{S(\tau + t)}{S(\tau)} - \mu t) ^2 \right)^{\frac{1}{1-q}}
\end{equation}
with  $\tilde{\beta}   = \beta(t)/ \sigma ^2 $.

This implies that the distribution of log-returns $\ln S(\tau + t)/S(\tau)$ 
over the interval $ t$ follows a Tsallis distribution, evolving  anomalously across timescales.
This result is consistent with  empirical 
evidence, in particular results found in  \cite{Michael&Johnson} for the S \& P 500. 
Consequently, the way in which the stochastic equation Eq(\ref{eq:lancondY}) with Eq(\ref{eq:nomega})  should be
 interpreted is that it generates members $Y(t)$ of an ensemble 
of returns, distributed  on each timescale $t$ according to a non-Gaussian Tsallis distributionof index $q$.   
With such an
interpretation, the  current model should be applicable to pricing both standard and exotic
 options, except for such options which are explicitly dependent on the history of evolution
 of a particular price path. However, in the current paper we only look at standard options.
 Exotics will be the topic of future study.  

At each timescale the distribution is of Tsallis form of index $q$. 
This appears to be true empirically,  based on studies in 
\cite{Michael&Johnson,osorio}, although for large $t$, empirical distributions do
seem to become Gaussian. However, this is consistent with our model because 
as $t$ increases, the central region of the Tsallis distribution is very well approximated 
by a Gaussian. Seeing as empirical data becomes sparser and sparser for large $t$, 
it is virtually impossible to say whether real returns become more and more Gaussian as the timescale increases, or whether they are still of Tsallis index $q$ and  only appear  
Gaussian due to a lack of empirical measurement in the tail region \cite{osorio}.

Another point which should be  addressed is the fact that the
 noise distribution at each timescale evolves according to a Tsallis distribution,
 with variance scaling anomalously with timescale. These results are consistent with 
empirical observations on short timescales \cite{Michael&Johnson} but it is more commonly 
found 
that the variance of returns scales normally at larger timescales.
Such scaling of the variance can be achieved with the current model simply by including
a state dependent instantaneous rate of return, for example, if the log returns show some form
of mean reversion. As shown in \cite{pdependent,Tsallis&Bukman}, 
the nonlinear Fokker-Planck equation corresponding to  such a problem  yields  as solutions Tsallis distributions
of index $q$ on each timescale, but  the exponent of the temporal
 evolution of $ <Y(t)^2> \propto 1/{\beta(t)}$  can be normal or even subdiffusive,
depending on both $q$ and the strength of the  mean reverting term.

However, for the purpose of asset pricing using Martingale techniques, the explicit properties of the deterministic part of the dynamics of the log returns become 
irrelevant, as long as the Novikov condition Eq(\ref{eq:crit}) is satisfied. We shall leave further 
exploration of these details for future work, but make a note that the results in this paper
could probably be extended to be valid for certian classes of models with 
state-dependent $\mu$.

\section{Risk-Free portfolio and the Generalized Black-Scholes Differential Equation}

Our model for log-returns reads  
\begin{equation}
\label{eq:langenlnS}
d Y = \mu dt  + \sigma  d \Omega
\end{equation}
with $Y(t) = \ln S(\tau+ t)/S(\tau)$ and
$d \Omega$ given by Eq(\ref{eq:nomega}).
In the following we shall set $\tau = 0$ without loss of generality for our
current discussion.
The stock price itself follows
\begin{equation}
\label{eq:langen1}
d S = (\mu + \frac{\sigma^2}{2}P_q^{1-q}) Sdt  + \sigma S d \Omega
\end{equation}
which can be abreviated  as
\begin{equation}
\label{eq:langenmu}
d S = \tilde{\mu} Sdt  + \sigma S d \Omega
\end{equation}
where 
\begin{equation}
\label{eq:tildemu}
\tilde{\mu} = \mu + \frac{\sigma^2}{2} P_q^{1-q}
\end{equation}
Remember that $P_q$ (given by Eq(\ref{eq:ptsalliscondzeromean})) is a function of $\Omega(t)$, so
 $\tilde{\mu}$ 
itself ultimately varies with time. (Having a time dependent rate
 of return is a perfectly valid assumption, even in the standard case).
The term $\frac{\sigma^2}{2} P_q^{1-q}$
 which appears here is none other than a noise-induced drift term.
For $q = 1$ the standard noise-induced drift term is recovered. This 
stock-price model 
implies that log returns are distributed according to the Tsallis 
distribution of Eq(\ref{eq:ptsallisnonzero}). 
(Note that a fully equivalent  treatment of the problem is to assume that
the dynamics of  the stock price is instead given by Eq(\ref{eq:langenS1}), as discussed in Appendix A).

Let us now look at price movements of a derivative of the underlying stock $S$, modelled by
Eq(\ref{eq:langenmu}).
We denote the price of the derivative by $f(S)$ and we use the stochastic
 (Ito) calculus to obtain
\begin{equation}
df = \frac{df}{dS}dS + \frac{df}{dt} dt + \frac{1}{2}\frac{d^2 f}{dS^2}(\sigma^2 P_q^{1-q}) dt
\end{equation}
where in turn $dS$ is given by   
Eq(\ref{eq:langenmu}) with Eq(\ref{eq:nomega}). After  insertion we get 
\begin{equation}
df = \left( \frac{df}{dS} \tilde{\mu} S + \frac{df}{dt} + \frac{1}{2} \frac{d^2f}{dS^2}(\sigma^2 S^2 P_q^{1-q})\right)dt + \frac{df}{dS}\sigma SP_q^{\frac{1-q}{2}} d\omega
\end{equation}
In the limit $q \rightarrow 1$, we recover the standard equations for price
 movements and derivatives thereof.

It is important to realize that the noise terms driving the price of the 
shares $S$ is the same as that driving the price $f$ of the derivative. 
It should be possible to invest one's wealth in a portfolio of shares and 
derivatives in such a way that the noise terms cancel each other, yielding 
the so-called risk-free portfolio. Following the same steps as in the standard
case (cf \cite{Hull}\cite{Neftci}),
\begin{equation}
\label{eq:pi}
\Pi = -f + \frac{df}{dS}S
\end{equation}
A small change in this portfolio is  given by 
\begin{equation}
\label{eq:deltapi}
\Delta \Pi = -\Delta f + \frac{df}{dS} \Delta S
\end{equation}
which, after insertion of the expressions for f and S, becomes
\begin{equation}
\Delta \Pi = - \left( \frac{df}{dt} + \frac{1}{2} \frac{d^2f}{dS^2} \sigma^2 S^2 P_q^{1-q}\right) \Delta t
\end{equation}
The return on this portfolio must be the risk-free rate r, otherwise there
 would be arbitrage opportunities. One thus gets the following generalized 
version of the Black-Scholes differential equation:
\begin{equation}
\frac{df}{dt} + \frac{1}{2} \frac{d^2f}{dS^2} \sigma^2 S^2 P_q^{1-q} 
= r(f - \frac{df}{dS} S)
\end{equation}
or rather
\begin{equation}
\label{eq:bsgen}
\frac{df}{dt}  + rS \frac{df}{dS}  + \frac{1}{2} \frac{d^2f}{dS^2} \sigma^2 
S^2 P_q^{1-q} 
= rf  
\end{equation}
where $P_q$  evolves according to Eq(\ref{eq:ptsalliscondzeromean}).
In the limit $q \rightarrow 1$, we recover the standard Black-Scholes
differential equation. 

This differential equation does not explicitly depend on ${\mu}$, 
the rate 
of return of the stock, only on the risk-free rate and the variance. 
However, there is a dependency on $\Omega(t)$ through the term $P_q$.
But it is possible to express $\Omega (t)$
in terms of $S(t)$ through Eq(\ref{eq:omtolns}), which implies  that there is
an implicit dependency on $\mu$. 
 Therefore, to be consistent with
 risk-free pricing theory, we should first transform our original stochastic 
equation for $S$
 into a  martingale before we apply the above analysis. This will
 not affect our results 
other than that $\tilde{\mu}$ will be replaced by the risk-free rate $r$. In the next Section we show how this can be done.

\section{Equivalent Martingale Measures} 

Assume that there is a call option with strike price $K$ written on the underlying asset $S(t)$. 
Its value will be given by
\begin{equation}
C(T) = \max[ S(T) - K, 0]
\end{equation}
at expiration date T. At earlier  times $ t < T$, the value of $C(T)$ is unknown
 but one can forecast it using the information $I(t)$  available up until 
time $t$, so
\begin{equation}
\label{eq:CT}
E^F[C(T) \mid I(t)] = E^F[\max[S(T) - K,0] \mid I(t)]
\end{equation}
where the notation $E^F[C]$ means that the expectation $E$ of the random
 variable $C$ is taken with respect to the probability measure $F$ under which the 
dynamics of $C$  (and thereby $S$) are defined. 
In addition, we
 must require that the fair market
value $C(t)$, discounted accordingly in the risk-neutral framework 
 at the risk-free rate, is 
equal to $E^F[\max[S(T) - K,0] \mid I(t)]$. However, this is only true if 
$e^{-rt}S(t)$
 satisfies the  martingale condition
\begin{equation}
E^F[e^{-rt}S(t) \mid S(u), u< t] = e^{-ru}S(u)
\end{equation}
This means that under the measure $F$, the conditional expectation of $S(t)$ 
discounted at the risk-free rate is best given by the discounted value of 
$S$ at the previous time $u$. Heuristically one can say that  a martingale is a 
stochastic process 
whose  trajectories display no obvious trends or periodicities. A 
submartingale is a process that, on average, is increasing.  
For example, using  the stock price  model of Eq(\ref{eq:langenmu}), we get  for $G(t) = e^{-rt}S(t)$
\begin{equation}
dG =( \tilde{\mu} - r  ) G dt + G\sigma  d \omega
\end{equation}
Clearly, $G$ is a submartingale because of the non-zero drift term, whereas 
the process 
\begin{equation}
dG - ( \tilde{\mu} - r ) G dt =  G\sigma d \omega
\end{equation}
is a martingale. Subtracting the drift from a submartingale $G$ in a somewhat similar manner is
 the basis of the so-called Doob-Meyer decomposition. If the drift term can be
explicitly determined, then it is possible to decompose $G$ into a drift 
component and a martingale component and thereby determine the fair market 
value of $C(t)$

However, this method is not usually used. Instead, 
it is common in asset-pricing to find synthetic probabilities $Q$ under
 which the drift of the underlying stochastic process vanishes, i.e., find $Q$ so that 
\begin{equation}
E^Q[e^{-rt}S(t) \mid S(u), u< t] = e^{-ru}S(u)
\end{equation}
In order to transform our probability-dependent
 stochastic processes into martingales, we will need to generalize several of
 the concepts used in the standard asset-pricing theory. Therefore, we shall
 first review the standard case.

If a stochastic process is given by
\begin{equation}
dY = \mu dt + \sigma d\omega
\end{equation}
where $\omega$ is a Brownian noise term associated with a probability 
measure $F$,
then it is not a martingale because of the drift term $\mu dt$. According to 
the Girsanov theorem, one can
 however find an equivalent measure $Q$ corresponding to an alternative 
noise term $dz$, such that the process is transformed into a martingale, by 
rewriting it as
\begin{equation}
dY = \sigma \left( \frac{\mu}{\sigma} dt + d \omega \right) = \sigma dz
\end{equation}
The new driving noise term $z$ is related to $\omega$ through 
\begin{equation}
z = \int_{0}^{t} u ds + \omega
\end{equation}
with 
\begin{equation}
u = \frac{\mu}{\sigma}
\end{equation}
The noise term $z$ is defined with respect to the 
equivalent Martingale  measure $Q$ which is related to $F$ through the 
Radon-Nikodym derivative
\begin{equation}
\label{eq:zeta}
\zeta(t) = \frac{dQ}{dF} = \exp\left( -\int_0^t u d\omega - \frac{1}{2} \int_0^t u^2 ds \right)
\end{equation}

Under the measure $F$, the original random variable $\omega$ follows a zero-mean process with variance equal to $t$. 
Under that same measure, the new noise term $z(t)$ is normal with non-zero 
mean equal to $\int_0^t u ds$ and variance $t$. However, with respect to the 
equivalent probability measure $Q$ 
one can easily verify that $z(t)$ is normal with 0 mean and variance $t$. This
follows because the relationship  
\begin{equation}
\label{eq:eqf}
E^Q[Y] = E^F[\zeta Y]
\end{equation}
holds. 
In  
the above discussion, $u$, $\mu$ and $\sigma$ may all depend on the variable
 $Y(t)$ as well. The only criterion which must be satisfied for the Girsanov theorem to be valid is that
\begin{equation}
\label{eq:crit}
\exp \left( - \frac{1}{2} \int_0^t u^2 ds \right) < \infty
\end{equation}
which implies that  $\zeta$ is a square integral martingale. This is known as the Novikov condition (for details see Oksendal \cite{Oksendal}).

The  effect of the martingale 
transformation is further illustrated by  
 the conditional probability distribution of the variable $Y$: With respect to $\omega$, $P$ is
given by
\begin{equation}
P(Y,t\mid Y(t_0),t_0)  = \frac{1}{ \sigma \sqrt{2\pi (t-t_0)}} \exp(\frac{-((Y-Y(t_0)) - \mu (t-t_0))^2}{\sigma^2 (t -t_0)})
\end{equation}
which is Gaussian with  drift $\mu (t -t_0)$.
 On the other hand, with respect to 
$z$, the probability distribution of $Y$  is 
given by 
\begin{equation}
P(Y,t\mid Y(t_0),t_0) = \frac{1}{\sigma \sqrt{2 \pi (t-t_0)}} \exp(\frac{-(Y-Y(t_0))^2}{\sigma^2 (t-t_0)})
\end{equation}
 This is a Gaussian distribution with  zero drift.

Now we would like to formulate similiar equivalent martingale  measures for 
the present class of probability dependent stochastic processes. 
Let the original process be given by
\begin{equation}
\label{eq:original}
dY = \mu dt + \sigma d \Omega
\end{equation}
with $\Omega$ defined as in Eq(\ref{eq:nomega}), namely
\begin{equation}
d \Omega = P_q^{\frac{1-q}{2}} d \omega
\end{equation}
where $\omega$ is normally distributed $\delta$-correlated noise, 
associated with the measure $F$.   $P_q(\Omega)$ is the 
Tsallis distribution of index $q$ discussed above Eq(\ref{eq:ptsalliscondzeromean}), so with respect to 
the measure $F$,  $P(Y,t \mid Y(t_0), t_0)$ is given by  
the non-zero drift distribution  
\begin{equation}
P_q(Y, t \mid Y(t_0), t_0) = \frac{1}{Z(t)} \left(1 - \tilde{\beta}(t) (1-q)( Y - Y(t_0)  -\mu (t - t_0)) ^2 \right)^{\frac{1}{1-q}}
\end{equation}

We are now in the position to define equivalent Martingale measures 
exactly as in the 
standard case by writing
\begin{eqnarray}
dY & = & \sigma P_q^{\frac{1-q}{2}} (\frac{\mu}{\sigma P_q^{\frac{1-q}{2}}} dt + d \omega)\\
   & = & \sigma P_q^{\frac{1-q}{2}} dz \label{eq:dYz}
\end{eqnarray}
This new driving noise $z$ is associated with the measure $Q$ and reads
\begin{equation}
\label{eq:dz}
dz = \frac{\mu}{\sigma P_q^{\frac{1-q}{2}}} dt + d \omega
\end{equation}
Let us  define
\begin{equation}
\label{eq:u1}
u = \frac{\mu}{\sigma P_q^{\frac{1-q}{2}}}
\end{equation}
Since $P_q$ is simply a particular function of $\Omega$, which in turn can be expressed as 
 a function of $Y$ via   $\Omega = (Y - \mu t) / \sigma$, we are dealing
with a general function $u(Y)$, so our analysis will be formally equivalent to that
 of the standard case. 
In particular, since $P_q$ is a non-zero bounded  function of $Y$  
the criterion 
Eq(\ref{eq:crit}) is  valid and thereby also the Girsanov theorem.
The martingale equivalent measure  $Q$ under which
$z$ is defined is given by Eq (\ref{eq:zeta}) with $u$ as in Eq(\ref{eq:u1}).
Under $Q$, the noise term $z$ is a zero-mean Brownian motion.
Remember that $z$ and $Q$ are merely synthetic measures. They are purely 
mathematical constructions that do not reflect the true probailities or
 dynamics of $Y$.   

The most important point that we shall utilize in this work is the following.
Since $z$ is under $Q$ a zero-mean Gaussian noise, then the noise term  defined by
\begin{equation}
d \Omega =  P_q(\Omega)^{\frac{1-q}{2}} dz 
\end{equation}
is equivalent to that defined by Eq(\ref{eq:nomega}) aqnd  the distribution of the variable 
$\Omega$ is therefore given by  a Tsallis distribution of index $q$.  Consequently, under $Q$, the variable $Y$ as defined by the stochastic 
equation Eq(\ref{eq:dYz}) is also distributed according to a zero-drift Tsallis distribution,
namely
\begin{equation}
P_q(Y, t \mid Y(t_0), t_0) = \frac{1}{Z(t)} \left(1 - \tilde{\beta}(t) (1-q)( Y - Y(t_0) ) ^2 \right)^{\frac{1}{1-q}}
\end{equation}

\section{Transforming the Discounted Stock Price to a Martingale}

In the following, we will discuss the problem of how to transform the
 discounted stock price into a martingale. Let the discounted stock price be
\begin{equation}
\label{eq:disc}
G = S e^{-rt}
\end{equation}
such that 
\begin{equation}
\ln G = \ln S - rt.
\end{equation}
The model for $S$ is given by Eq(\ref{eq:langenmu}),  
 yielding 
\begin{equation}
\label{eq:dg}
dG = ( \tilde{\mu} - r ) G dt + \sigma G d \Omega
\end{equation}
for the discounted stock price $G=Se^{-rt}$.
The dynamics of $\Omega$ is defined with respect to the measure $F$ as in Eq (\ref{eq:nomega}).
 Here, $\tilde{\mu}$ includes a noise-induced drift term 
and reads as in Eq(\ref{eq:tildemu}).
Stochastic integration  shows that at time $T$ we have 
\begin{eqnarray}
G(T) &= &G(0) \exp ( (\tilde{\mu} - r)T
 - \frac{\sigma^2}{2}\int_0^T P_q^{1-q}dt
 + \int_0^T \sigma P_q^{\frac{1-q}{2}} dz_s  )\\
&= &G(0) \exp ( (\mu - r)T
+  \int_0^T \sigma P_q^{\frac{1-q}{2}} dz_s )
\end{eqnarray}
which implies
\begin{eqnarray}
\label{eq:SF}
S(T)& = & S(0) \exp \left( \int_0^T \sigma P_q^\frac{1-q}{2}d \omega_s  + \int_0^T(\tilde{ \mu}    - \frac{\sigma^2}{2} P_q^{1-q})dt \right)\\
& = & S(0) \exp \left( \int_0^T \sigma P_q^\frac{1-q}{2}d \omega_s  +
 \mu T  \right)
\end{eqnarray}
These expressions are derived based on the original representation of the price dynamics, given by
Eq(\ref{eq:dg}). Hovwever it is clear that Eq(\ref{eq:dg})  is not
a martingale, but can be transformed into one following the same ideas  as discussed in the previous 
Section. We get
\begin{equation}
\label{eq:dgz}
dG =  \sigma G P_q^{\frac{1-q}{2}} dz
\end{equation} 
with 
\begin{eqnarray}
dz &= &(\frac{ \tilde{\mu} -r }{
\sigma P_q^{\frac{1-q}{2}}})dt +
 d \omega\\
 &= &(\frac{ \mu -r +\frac{\sigma ^2 P_q^{1-q}}{2}}{\sigma P_q^{\frac{1-q}{2}}})dt + d \omega\\
\end{eqnarray}
Notice that $P_q$ depends on $\Omega$ which in turn depends on $S$ as was
 shown in Eq(\ref{eq:omtolns}). $S$ itself can be expressed  in terms of
 $G$ via Eq(\ref{eq:disc}). Therefore, the rules of standard stochastic 
calculus can be applied, and 
the martingale equivalent measure $Q$ associated with $z$ is obtained 
from Eq(\ref{eq:zeta}) by setting 
\begin{equation}
u = \frac{ \mu - r  + \frac{\sigma ^2 P_q^{1-q}}{2}}{\sigma P_q^{\frac{1-q}{2}}}
\end{equation}
Taking the log of Eq(\ref{eq:dgz}) we get
\begin{equation}
\label{eq:dlngz}
d \ln G =  -\frac{\sigma^2}{2}P_q^{1-q} dt  + \sigma P_q^{\frac{1-q}{2}} dz
\end{equation} 
After stochastic integration and transforming back to $S$ we obtain
\begin{eqnarray}
\label{eq:mstock}
S(T)& =& S(0) \exp \left( \int_0^T \sigma P_q^{\frac{1-q}{2}} dz_s 
+ \int_0^T (r - \frac{\sigma^2}{2} P_q^{1-q}) ds \right)  \\
%& = & S(0) \exp \left( \int_0^T \sigma P_q^{\frac{1-q}{2}} dz_s + r T - \frac{\sigma^2 }{2}( \alpha T^{\frac{2}{3-q}}  + (1-q)\int_0^T \frac{\beta(t) }{Z(t)^{1-q}}\Omega^2(t) dt)    \right) \nonumber\\
%& = & S(0) \exp \left( \Omega(T) + r T - \frac{\sigma^2 }{2} \alpha T^{\frac{2}{3-q}}  + (1-q)\frac{\sigma^2}{2}\int_0^T \frac{\beta(t)}{Z(t)^{1-q}} \Omega^2(t) dt    \right) \nonumber
\end{eqnarray}
with
\begin{equation}
\label{eq:alpha}
\alpha =  \frac{1}{2}(3-q) ((2-q)(3-q)c)^{\frac{q-1}{3-q}}
\end{equation}
If we compare the expression Eq(\ref{eq:mstock}) for $S$   under $Q$ with that 
under F as given by  Eq(\ref{eq:SF}), we see that the difference between the two is that 
the rate of return $\tilde{\mu}$ has been replaced by the risk-free rate $r$. 
This 
recovers the same result as in the standard risk-free asset pricing theory.
(Exactly the same result
Eq(\ref{eq:mstock}) would have been obtained had we instead started with the stock price model Eq(\ref{eq:langenS1}), as mentioned in Appendix A. It is not hard to see that 
that  would have been 
equivalent to substituting $\mu$ with the risk free rate $r$).

We have yet to discuss the evaluation of the $P_q$ related terms which appear in the above expressions.
Two points will be of importance here: The first being that the term of type
 $\int_0^T P_q^{\frac{1-q}{2}} dz $ is simply equal to the random variable $\Omega(T)$.
The second important point (discussed in Appendix B) is to realize that the distributions 
$P_q(\Omega(s))$ at arbitrary times $s$ can
be mapped onto the distributions $P_q(\Omega(T))$ at a fixed time $T$ via the appropriate variable 
transformation 
\begin{equation}
\Omega(s)) = \sqrt{\frac{\beta(T)}{\beta(s)}} \Omega(T)
\end{equation}
Using these notions we can  write $S(T)$ of Eq(\ref{eq:mstock}) as
\begin{eqnarray}
S(T)& = & S(0) \exp \left( \Omega(T) + r T - \frac{\sigma^2 }{2} \alpha T^{\frac{2}{3-q}}  + (1-q)\frac{\sigma^2}{2}\int_0^T \frac{\beta(t)}{Z(t)^{1-q}} \Omega^2(t) dt    \right) \nonumber  \\
&=&S(0) \exp \left( \Omega(T) + r T - \frac{ \sigma^2 }{2}\alpha T^{\frac{2}{3-q}}( 1 - (1-q) \beta(T) \Omega^2(T) )    \right) \label{eq:stomega}
\end{eqnarray}
This expression for $S(T)$ recovers the usual one for $q=1$. For $q >1$, 
a major difference to the standard case is the $\Omega^2(T)$-term in the exponential,
 which appears as a result of the noise induced drift. The implications of this term
for the option prices will become apparent  further on.

Let us  revisit  the generalized Black-Scholes PDE
Eq(\ref{eq:bsgen}). In the risk-neutral world, we must use Eq(\ref{eq:dlngz}) to obtain an expression for $P_q(\Omega)$. Rewriting that equation yields
\begin{equation}
\label{eq:dlnsz}
d \frac{(\ln S - rt  + \frac{\sigma^2}{2} P_q^{1-q}(\Omega)  )}{\sigma} =  d\Omega
\end{equation} 
Formally, this expression is identical to Eq(\ref{eq:omtolns}) except that $\tilde{\mu}$ (related to $\mu$ through Eq(\ref{eq:tildemu})) has been replaced with $r$. 
Furthermore, integrating Eq(\ref{eq:dlnsz}) up to time $t$ results in  Eq(\ref{eq:stomega}) (with $T= t$), 
from which it is possible to solve for $\Omega(t)$ explicitly in terms of $S(t)$. 
This implies that, in the martingale representation,  $P_q(\Omega(t))$ can be expressed as
a function of  the volatility $\sigma$, the risk-free rate $r$, $S(t)$ and $S(0)$. Most importantly,
the implicit dependency on $\mu$ through $\Omega$ is replaced by a dependency on $r$. 

The generalized differential equation Eq(\ref{eq:bsgen}) can thus  be solved numerically,
which is one way of obtaining option prices in this generalized framework. 
However,  it is possible to go a step further and obtain 
closed-form option
 prices. This is done  by transforming asset prices into martingales and
then taking expectations. In the following Sections we show how this is done,
 and  why the option prices 
obtained in that way indeed satisfy Eq(\ref{eq:bsgen}).

\section{The Generalized Option Pricing Formula}

Suppose that we have a European claim $C$ which depends on $S(t)$, whose 
price $f$  is given by its expectation value in a risk-free (martingale)
 world  as
\begin{equation}
\label{eq:fgenc}
f(C) = E^Q [e^{-rT} C]
\end{equation}
If the payoff on this option depends on the stock price at the expiration time $T$ so that
\begin{equation}
C = h(S(T))
\end{equation}
then we obtain
\begin{equation}
\label{eq:fprice1}
f = e^{-rT} E^Q\left[h \left( S(0) \exp \left( \int_0^T \sigma P_q^{\frac{1-q}{2}} dz_s + \int_0^T(r - \frac{\sigma^2}{2} P_q^{1-q}) ds \right) \right) \right]
\end{equation}
In the special case of $q = 1$, the standard expression of the option price is recovered with this formula (see for example Oksendal \cite{Oksendal}). However, in that case it is argued that under $Q$, the random variable
\begin{equation}
x(T) = \int_0^T \sigma dz
\end{equation} 
is normally distributed with variance
\begin{equation}
\delta^2 = \int_0^T \sigma^2 dt
\end{equation}
yielding the following expression for a European claim:
\begin{equation}
f = \frac{e^{-rT}}{\delta \sqrt{2 \pi}} \int_R h \left[ S(0) \exp\left( x + \int_0^T (r - \frac{1}{2}\sigma^2(s) ) ds \right) \right]
\exp\left( - \frac{x^2}{2 \delta ^2} \right) dx
\end{equation}
The key difference in our  approach is that the
 random variable   
\begin{equation}
\label{eq:dxx1}
\frac{x(T)}{\sigma}   = \int_0^T  P_q^{\frac{1-q}{2}} dz_s = \Omega(T)
\end{equation}
is not normally distributed, but rather 
according to the Tsallis distribution of index 
$q$ Eq(\ref{eq:ptsalliscondzeromean}). 
The pricing equation Eq(\ref{eq:fprice1}) can be 
written as 
\begin{eqnarray}
f &=& \frac{e^{-rT}}{Z(T)} \int_R h\left[ S(0)\exp(\sigma \Omega(T) +rT - \frac{\sigma^2}{2} \alpha T^{\frac{2}{3-q}}(1  
 - (1-q) \beta(T) \Omega^2(T)) )  \right]
\nonumber \\ & & (1-\beta(T)(1-q)\Omega(T)^2)^{\frac{1}{1-q}} d\Omega_T
\end{eqnarray}
In the limit $q = 1$, the standard result is recovered.

\section{European Call Options}
A European call option is such that the option holder has the right to 
buy  the  underlying stock $S$ at the strike price $K$, on the day of expiration 
$T$. Depending on the value of $S(T)$, the payoff of such an option is
\begin{equation}
C = \max [ S(T) - K, 0]
\end{equation}
In other words, if  $S(T) > K$ then the option will have value 
(it will be in-the-money). In a  more concise notation, the price $c$ of such 
an option becomes
\begin{eqnarray}
\label{eq:optionprice}
c & = & E^Q [e^{-rT} C]\\
 & = & E^Q [ e^{-rT} S(T)]_D - E^Q [ e^{rT} K]_D\\
& = & J_1 - J_2
\end{eqnarray}
where the subscript $D$ stands for the set $\{ S(T) > K \}$. To calculate $J_1$ and $J_2$ we shall proceed along the same lines as in the standard case \cite{Marek&Marek}. We have
\begin{eqnarray}
J_2 & = &  e^{-rT}K \left( \int_{R}\frac{1}{Z(T)}(1- \beta(T)(1-q)\Omega(T)^2)^{\frac{1}{1-q}} d \Omega_T\right) _D\\ 
  & = & e^{-rT} K {\bf P_Q} \{ S(T) > K \}
\end{eqnarray}
where the notation ${\bf P_Q} \{ S(T) > K \} $ is just a more concise notation for
the expression on the line above. ${\bf P_Q }$ corresponds to the integral over the
 Tsallis distribution (which was defined with respect to the measure $Q$),
 and the argument $\{ S(T) > K \}$ is referring to the fact that
we are considering only the set $D$.
We get 
\begin{eqnarray}
J_2 & = &  e^{-rT}K {\bf  P_Q} \{ S(T) > K \}  \label{eq:jj2}  \\
&= & e^{-rT} K {\bf P_Q }\{ S(0) \exp ( \sigma \Omega + rT - \frac{\sigma^2}{2}\alpha T^{\frac{2}{3-q}}
(1  - (1-q)\beta(T)\Omega^2)  ) > K \}
\nonumber  \\
&= & e^{-rT} K {\bf P_Q }\{ -\frac{\sigma^2}{2} \alpha T^{\frac{2}{3-q}}( 1 -
(1-q) \beta(T) \Omega^2) + \sigma \Omega  + rT > \ln \frac{K}{S(0)}
\} \nonumber 
\end{eqnarray}
The inequality
\begin{equation}
-\frac{\sigma^2}{2} \alpha T^{\frac{2}{3-q}} +
(1-q)\alpha T^{\frac{2}{3-q}} \beta(T)\frac{\sigma^2}{2} \Omega^2 + \sigma \Omega + rT > \ln \frac{K}{S(0)} 
\end{equation}
is satisfied inbetween the two roots
\begin{eqnarray}
\label{eq:roots}
s_{1,2} & = &  \frac{-1}{\alpha T^{\frac{2}{3-q}} (1-q)\sigma \beta(T)} \\
&\pm& [\frac{1}{ \alpha T^{\frac{4}{3-q}} (1-q)^2\sigma^2 \beta(T)^2}\nonumber \\& -&
  \frac{2 }{ (1-q)  \alpha T^{\frac{2}{3-q}}\sigma^2 \beta(T)}(rT + \ln\frac{S(0)}{K} - 
\frac{\sigma^2}{2} \alpha T^{\frac{2}{3-q}}) ]^{\frac{1}{2}} \nonumber
\end{eqnarray}

This is a very different situation from the standard case, where the inequality
is linear and the condition $S(T) > K$ is satisfied for all values of the
random variable greater than a threshold. In our case, due to the noise induced
 drift, values of $S(T) $ in the risk-neutral world are not monotonically 
increasing as a function of the noise.  As $q \rightarrow 1$, the larger 
root goes toward $\infty$, recovering the standard case. But as $q$ gets 
larger, the tails of the noise distribution get larger, as does the noise induced drift which tends to pull the system back. 
As a result we obtain
\begin{eqnarray}
J_2 & = & \frac{e^{-rT} K}{Z(T)} \int_{s_1}^{s_2} (1 - (1-q) \beta(T) \Omega^2)^{\frac{1}{1-q}} d\Omega 
\end{eqnarray}

The remaining term $J_1$ can be determined in a similar fashion. We have
\begin{equation}
\label{eq:j1}
J_1 = E^Q [ e^{-rT}S(T) ]_D
\end{equation}
This can be written as
\begin{eqnarray}
J_1 & =& E^Q [e^{-rT}S(T) ]_D \\
& = & {\bf P_Q } [e^{-rT}S(T)] \{ S(T) > K \} \nonumber \\
& = & {\bf P_Q } [e^{-rT} \exp ( S(0)\exp(\sigma \Omega(T) +rT - \frac{\sigma^2}{2}  \alpha T^{\frac{2}{3-q}}(1
 - (1-q) \beta(T) \Omega^2(T)) ) ]\nonumber \\
& & \{ S(T) > K \} \nonumber
\end{eqnarray}
The domain $\{ S(T) > K \}$ is the same 
as that found for $J_2$, and is
defined as the region between the two roots of Eq(\ref{eq:roots}).
We obtain
\begin{eqnarray}
J_1 
&  = &  \frac{S(0)}{Z(T)} \int_{s_1}^{s_2} \exp( \sigma \Omega -
\frac{\sigma^2}{2} \alpha T^{\frac{2}{3-q}} 
 + (1-q) \alpha T^{\frac{2}{3-q}} \beta(T) \frac{\sigma^2}{2} \Omega^2 ) \nonumber \\
& & (1 - (1-q)  \beta(T) \Omega^2)^{\frac{1}{1-q}} d\Omega \label{eq:J1} 
\end{eqnarray}

It is customary in the standard Black-Scholes case  to express 
the integrals in Eq(\ref{eq:J1}) and Eq(\ref{eq:jj2})  in terms of  a 
standardized (0,1) noise process. It is possible to do the same in the
generalized case, via the apropriate variable transformation
\begin{equation}
\Omega_N = \Omega(T)\sqrt{\frac{\beta(T)}{\beta_N}}
\end{equation}
We thus obtain the following expression for a European call option: 
\begin{equation}
\label{eq:call}
c = S(0)M_q(d_1,d_2, b(\Omega_N)) - e^{-rT}KN_q(d_1,d_2)
\end{equation}
where we introduce the notation
\begin{equation}
\label{eq:ndd}
N_q(d_1,d_2)   = \frac{1} {Z_N} \int_{d_1}^{d_2} (1 - (1-q)\beta_N  \Omega_N^2)^{\frac{1}{1-q}} d\Omega_N
\end{equation}
and
\begin{equation}
\label{eq:mdd}
M_q(d_1,d_2,b(\Omega_N)) = \frac{1}{Z_N}\int_{d_1}^{d_2} \exp( b(\Omega_N))(1 -(1-q)\beta_N \Omega_N^2)^{\frac{1}{1-q}} d\Omega_N
\end{equation}
with 
\begin{equation}
b(\Omega_N) = \sigma \sqrt{\frac{\beta_N}{\beta(T)}} \Omega_N - \frac{\sigma^2}{2}\alpha T^{\frac{2}{3-q}} ( 1 - (1-q) \beta_N \Omega_N^2)
\end{equation}
The  limits of the standardized integrals  are given as
\begin{equation}
d_{1,2} = \frac{s_{1,2}} {\sigma \sqrt{\beta_N/\beta(T)}}
\end{equation}
with $s_{1,2}$ as in Eq(\ref{eq:roots}).
By choosing  $\beta_N$ as 
\begin{equation}
\label{eq:betan}
\beta_N = \frac{1}{5-3q}
\end{equation}
the variance of the noise distribution will be normalized to 1 for each value of $q$.
In  the limit $q = 1$, the standard Black-Scholes pricing equations are 
recovered.

\section{Martingale Solutions and the Generalized Black-Scholes Differential Equation}

We must yet discuss the equivalence of the solution $f$ found via the martingale
asset pricing approach, and the solution of the generalized Black-Scholes differential
 equation (\ref{eq:bsgen}). We use arguments based on those in \cite{Shreve} for the
standard case.  The expression for $S$  of  Eq(\ref{eq:mstock}) can be written for
 $u  \ge t$ as
\begin{equation}
S(u) =  S(t) \exp \left( \int_t^u \sigma P_q^{\frac{1-q}{2}} dz_s +
\int_t^u (r - \frac{\sigma^2}{2} P_q^{1-q}) ds \right)
\end{equation}
This implies that  
\begin{eqnarray}
S(T) & = & S(0) \exp \left( \int_0^T \sigma P_q^{\frac{1-q}{2}} dz_s +
\int_0^T (r - \frac{\sigma^2}{2} P_q^{1-q}) ds \right)
\end{eqnarray}
can trivially be rewritten as
\begin{eqnarray}
S(T) & = & S(t) \exp \left( \int_t^T  \sigma P_q^{\frac{1-q}{2}} dz_s + 
\int_t^T (r - \frac{\sigma^2}{2} P_q^{1-q}) ds \right)\\
& =& XY
\end{eqnarray}
where
\begin{eqnarray}
X &=& S(t) \\
Y&  =& \exp \left( \int_t^T  \sigma P_q^{\frac{1-q}{2}} dz_s + 
\int_t^T (r - \frac{\sigma^2}{2} P_q^{1-q}) ds \right)
\end{eqnarray}
with the important properties that $X$ is measurable with information $I(t)$ available
up until time $t$, and $Y$ is independent of that information.

We then define
\begin{eqnarray}
v(t,X) & = &  E^Q[ h(S(T)) \mid I(t) ] \\
& = & E^Q [ h( X  \exp ( \int_t^T \sigma P_q^{\frac{1-q}{2}} dz_s +
\int_t^T (r - \frac{\sigma^2}{2} P_q^{1-q}) ds )
\end{eqnarray}
where $h$ is an arbitrary function.
We now look at the value of this expectation conditioned on information $I(t)$ 
available up until time $t$ and obtain 
\begin{eqnarray}
E^Q[ h(S(T)) \mid I(t)] & = &E^Q[h(XY)] \mid I(t) ]\\
& = & E^Q[h(X) \mid I(t)] \\
& = & v(t,X) \\
& = & v(t,S(t))
\end{eqnarray}
where the independence of $Y$ on $I(t)$ has been used.
 This is exactly the same result as obtained in the standard case,
and it  implies that $v(t,S(t)), 0 \le t \le T$, is a martingale \cite{Shreve}.
We proceed to use Ito's formula to write
\begin{equation}
d v(t, S(t)) = [ \frac{dv}{dt}  + r S \frac{dv}{dS} + \frac{1}{2} \sigma^2 S^2 P_q^{1-q} \frac{d^2v}{dS^2} ] dt 
+ \sigma S \frac{dv}{dS} P_q^{\frac{1-q}{2}} dz
\end{equation}
But because $v$ is a martingale, we know that the sum of the $dt$ terms must equal 0.
This implies that
\begin{equation}
\label{eq:fkac}
\frac{dv}{dt}  + r S \frac{dv}{dS} + \frac{1}{2} \sigma^2 S^2 P_q^{1-q} \frac{d^2v}{dS^2}  = 0
\end{equation}
for $ 0 \le t \le T$, which is  consistent with the Feynman-Kac theorem (cf \cite{Shreve,Oksendal}), albeit now generalized to the current framework. 

Recall that the price of a contingent claim paying $h(S(T))$ can be written as
 Eq(\ref{eq:fgenc}) so that
\begin{eqnarray}
f& = & E^Q [e^{-r(T-t)} C]\\
&  = & e^{-r(T-t)} E^Q[ h(S(T))] \\
&  = & e^{-r(T-t)} v(S,t)
\end{eqnarray}
implying  that
\begin{equation}
v(S,t) = e^{r(T-t)}f
\end{equation}
Insertion of this form of $v$ into Eq(\ref{eq:fkac})  immediately yields our
generalized Black-Scholes partial differential equation of Eq(\ref{eq:bsgen}).

We have thus shown that the option price $f$ obtained by way of transforming the
asset price into a martingale and discounting it accordingly (as represented by
 Eq(\ref{eq:fgenc})) in turn  implies that the generalized Black-Scholes equation 
of Eq(\ref{eq:bsgen}) must be valid.
Therefore, equivalent solutions can be found either by solving Eq(\ref{eq:bsgen}) or
Eq(\ref{eq:fgenc}).

\section{Dividends  and Futures }

We shall now show that the current model can also be generalized in a
 straightforward way to  give the price of options on dividend paying stocks, as well as
options on futures contracts of
the underlying stock. The futures markets are
 widely traded,  so being able to price these instruments within the current
 framework could be very useful. 

We first 
look at the case of a dividend paying stock.  Following standard arguments \cite{Hull}, in time $\Delta t$ 
the portfolio $\Delta \Pi$  (Eq (\ref{eq:pi})) gains wealth equal to 
$\Delta \Pi$ as in Eq (\ref{eq:deltapi})  as well as
dividends equal to
\begin{equation}
wS \frac{\partial f }{\partial S} \Delta t
\end{equation}
where $w$ denotes a continuous dividend yield.
The generalized Black-Scholes differntial equation thus becomes
\begin{equation}
\label{eq:bsgendiv}
\frac{df}{dt}  + (r-w)S \frac{df}{dS}  + \frac{1}{2} \frac{d^2f}{dS^2} \sigma^2 
S^2 P_q^{1-q} 
= rf  
\end{equation}

In the  risk-neutral martingale representation, this is equivalent to taking
the discounted expectation of a stock yielding a return $r-w$. For European 
calls we obtain
\begin{equation}
c =S(0) e ^{-wT}M_q(e_1,e_2, b(\Omega_N)) - e^{-rT}KN_q(e_1,e_2)]
\end{equation}
using the same notation as in Eq(\ref{eq:ndd}) and Eq(\ref{eq:mdd}), but
where $e_1$ and $e_2$ are solutions to
\begin{eqnarray}
\label{eq:eroots}
e_{1,2} & = &  \frac{-1}{\alpha T^{\frac{2}{3-q}} (1-q)\sigma \beta(T)} \\
&\pm& [\frac{1}{ \alpha T^{\frac{4}{3-q}} (1-q)^2\sigma^2 \beta(T)^2}\nonumber \\& -&
  \frac{2 }{ (1-q)  \alpha T^{\frac{2}{3-q}}\sigma^2 \beta(T)}((r-w)T + \ln\frac{S(0)}{K} - 
\frac{\sigma^2}{2} \alpha T^{\frac{2}{3-q}})]^{\frac{1}{2}} \nonumber
\end{eqnarray}

The evaluation of options on futures  is now straightforward,
since 
one argues that a futures contract $F$ is equivalent to a stock paying 
dividends $w$  exactly equal to the risk-free rate of return $r$. 
Therefore, for this case we obtain 
\begin{equation}
\label{eq:bsgenfut}
\frac{df}{dt}  \frac{1}{2} \frac{d^2f}{dF^2} \sigma^2 
F^2 P_q^{1-q} 
= rf  
\end{equation}
The closed form solution for European calls follows as
\begin{equation}
\label{eq:genblack}
c = e ^{-rT}[F(0) M_q(e_1,e_2, b(\Omega_N)) - KN_q(e_1,e_2)]
\end{equation} 
with $e_1$ and $e_2$ given as in Eq(\ref{eq:eroots}) with $S(0)$ substituted by $F(0)$ 
and  $w=r$.
For $q=1$, this  is known as the Black model.

\section{Numerical Results and The Greeks}

We evaluated European call options using Eq(\ref{eq:call}), and confirmed these
results  by numerically  solving Eq(\ref{eq:bsgen}) on a grid under appropriate boundary 
conditions. 
It is of particular interest to evaluate call options
 and see how the option prices
and partials change  as $q$  moves away from 1,
which recovers the   Black-Scholes scenario. 

Results of such calculations are shown in Figures 3 onward. Figure 3 depicts 
 the call option price
 as a function of the strike price for the standard Black-Scholes model ($q=1$)
 and our model with $q= 1.5$, where $\sigma$ is chosen such that the at-the-money 
 prices are equal. The differences between the two pricing models is more apparent in Figure 4.  There it is clear that both in-the-money and out-of-the money options
 are valued higher with $q=1.5$, except for very deep-in-the-money options which are valued lower.
This behaviour can be understood intuitively as follows. The distribution of $\Omega$ for $q=1.5$ has fatter tails
than the $q=1$ model.  
 Consequently, if the stock price gets 
deep out-of-the-money, then the noise may still produce shocks that can bring the stock back in-the-money again. 
This results in higher option prices for deep out-of-the-money strikes. Similarly, if the option  is deep in-the-money, the noise can
produce shocks to the underlying which can bring the price out-of-the-money again. In addition,  it can be seen from the expression
Eq(\ref{eq:mstock}) for $S(T)$, that large shocks will increase the value 
of the noise-induced drift term which will  decreases the probability of realizing  higher  stock prices.   This results in
lower option prices for deep-in-the-money strikes. On the other hand, for  intermediate values around-the-money, there will be a higher 
probability to land both in- or out-of-the-money which leads to an increase in the option price, relative to the standard $q=1$ model.

 The resulting volatilities which the  
standard model must assume in order to match the values obtained for the $q=1.5$ model, are plotted in Figure 5, for $T=0.1$ and $T=0.6$. Clearly, these implied volatilities (shown here for values $\pm 20 \%$ around-the-money)
 form a smile shape, very similar to that which is implied by real market data. The higher volatility $q=1$ Gaussian models
 that are successively needed as
one moves away-from-the money  essentially reflects the fact that the tails of the $q=1.5$ model would have to be 
approximated by higher volatility Gaussians, whereas the central part of the $q=1.5$ noise distrubution can be approximated by 
lower volatility Gaussians.

In Figure 6, the call option price as a function of time to expiration $T$ is plotted, for $q=1 $ and $q=1.5$. Figure 7 shows the call price as a function of the parameter $q$ for $T=0.4$. As $q$ increases, the three curves corresponding
to strikes  in-the-money, at-the-money, and out-of-the-money all behave similarily. 
However, the behaviour looks different for smaller T, as is seen in   Figure 8 where $T=0.05$. In Figure 9, the call option price as a function of $\sigma$ is shown, for $q=1 $ and $q=1.5$. In all of these plots, we use parameters close to those in \cite{Rubinstein}, where one can verify  our  results for $q=1$.

Figures 10 onward show the so-called Greeks as a function of the current
stock price. The Greeks are partial derivatives defined as 
\begin{eqnarray}
\Delta & = &\frac{\partial f}{\partial S}\\
\theta & = &-\frac{\partial f}{\partial T}\\
\kappa & = &\frac{\partial f}{\partial \sigma}\\
\rho & = &\frac{\partial f}{\partial r}\\
\Gamma & = &\frac{\partial \Delta}{\partial S} = \frac{\partial ^2 f}{\partial S^2}
\end{eqnarray}
In accordance, we introduce a new Greek designated by the symbol
Upsilon ($\Upsilon$), to represent the partial with respect to $q$, namely
\begin{eqnarray}
\Upsilon & = &\frac{\partial f}{\partial q}
\end{eqnarray}

\section{Empirical Results}

The real test of any model is  how well it can predict or describe 
empirical data. When it comes to option pricing, the standard Black-Scholes
formula misprices observed market prices in a rather systematic way. 
In particular, if for example $\sigma$ is chosen so that  the theoretical 
 at-the-money call price matches the market price, then the model will underprice  out-of-the money and in-the-money calls. Instead, to obtain theoretical prices which match the observed ones, a different value of $\sigma$ must be used
for each value of the strike. A plot of $\sigma$ versus the strike $K$ is typically a convex function, dubbed the volatility smile. The smile changes also with the time to expiration of the option, flattening out for larger times. A plot of $\sigma$ over $K$ and $T$  is known as the volatilty surface.
The fact that this surface is not constant is an indication that the option
 values predicted by the Black-Scholes model deviate from empirically 
observed ones.

To test our model, we shall use a value of $q= 1.4$, which is a good fit to the empirically observed returns distribution of financial data such as the S\&P 500. We shall then calibrate $\sigma$ so that the theoretical at-the-money 
 call matches market data, for a given time to expiration $T$. We shall then calculate option prices  using that one value of $\sigma$ across different strikes.
The next step is to find the different values of $\sigma$  which a standard Black-Scholes model would need in order to yield the same prices as our $q=1.4$ model. 
That will result in a volatility smile, which we can compare to the empirically observed volatilty smile.  If our model is a good one of market data, then
the smiles produced by the model should closely agree with the observed smiles.
Furthermore, this should hold true for many different times to expiration.

We have performed just such an experiment on options on Japanese Yen futures.
The market data of call prices is readily available, for example on \cite{bpfutures}.
Market smiles are backed out using a standard Black model (Eq(\ref{eq:genblack}) with $q=1$), and are plotted in Figures  16 and 17. The relevant values of $F(0)$, $r$ and $T$ are noted in the figure captions.
 We then used the generalized Black model Eq(\ref{eq:genblack}) with $q=1.4$ 
to obtain  theoretical call prices, using one value of $\sigma$ for each $T$,
chosen so that the theoretical at-the-money all price matched the market at-the-money call
(see
 Figures 16 and 17).  Volatility smiles implied by the $q=1.4$ model 
 were backed out using a standard Black model (q=1),
 and these are also plotted. One sees a very good agreement between market
 smiles and those implied from our model, for times to expiration ranging from
 17 days to 147 days. We would have included longer times to expiration but 
there was hardly any volume on those options.

We can also plot the at-the-money implied volatility as a function of time 
to expiration $T$. This is known as the volatility term structure, and is shown in Figure 18.
 Interestingly, the $q=1.4$ at-the-money volatility parameter $\sigma$ decreases with $T$, in
 roughly the same way as the at-the-money $\sigma$ increases with $T$ for the standard $q=1$
 model.
 Note however, that while the entire volatility surface consisting of the variation of
 $\sigma$ across strikes as well as across time is needed for the $q=1$ model to fit empirical
 data, the $q=1.4$ volatility surface is captured by  the  plot of $\sigma$ versus $T$ of  
Figure 18, seeing as it is that same value of $\sigma$ which is used across all strikes.

These results are encouraging. 
 In particular,  please note that we did not in any way optimize  our choice of $q=1.4$. It may will be that there is another  slightly different value of $q$ which  well models empirical smiles as well as
 produces a volatility which is constant with  respect to $T$.
Even so, for a practitioner using this model, a
 deterministic term structure in
 volatility can easily be hedged away for portfolio management purposes.

Finally, a few brief words comparing the reuslts of our model to other 
 models which have been introduced to accomadate the volatility smile.
For example the DVF (Deterministic Volatility Function) approach \cite{Dupire},
where the smile is a consequence of the volatility being time and state dependent. In principle, this is not entirely different to some ideas in 
our current approach, but  in the DVF model the volatility surface shows extreme variations for shorter times to expiration, and is highly
 nonstationary in general. 

Jump diffusion models \cite{Merton76} are another class of 
interesting stochastic processes which have been introduced to explain 
smiles. In such models, the volatility smile is explained by adding
discontinuous Poisson jumps to the standard Black-Scholes model of stock price movement.  These models are however highly parametrized and  difficult to handle numerically. 

Stochastic volatility models are yet another approach to modeling 
the volatility smile. Here the volatility is assumed to follow a stochastic process, which could be a GARCH process or a mean-reverting diffusion process  (cf \cite{Hull&White}). These models again assume more parameters, do not allow for closed-form solutions, and can be difficult to handle numerically. Combinations of some of the approaches listed here have also been studied, for example \cite{jesper}.

With this  said, a clear advantage of the approach which we present in this paper,
is that we can work with many of the useful tools of the standard Black-Scholes approach, such as  techniques from risk-free asset pricing.
Numerically, our generalized PDE is easy to solve, and perhaps above all, we obtain  
closed form solutions for certain special cases. A final point is that we need much fewer parameters to well-model empirically observed option prices,
 (just one value of $\sigma$ across all strikes).
Our model seems to work well for options on currencies, bonds and perhaps on single-stock options. Even though it also provided a good fit to options on the S \& P 100 index \cite{shortpaper},  in general the volatility smiles observed in such markets tend to have more of a skew, or a ``smirk''. Introducing an assymmetry into the noise distribution may be one way of extending our model to such scenarios.

\section{Conclusions}

In summary, we have proposed modelling the random noise affecting stock
returns  as evolving across timescales according to an anomalous Wiener 
process characterised by
a Tsallis distribution of index $q$. This non-Gaussian noise  satisfies a
statistical
feedback process which ultimately depends on a standard Brownian motion.
 We conclude that 
our approach  yields a better 
description than using standard normally distributed noise, because  
we obtain processes whose distributions match  
empirical ones  much more closely, while including the standard results 
 as a special case.
Based on these novel stochastic processes,  
 a generalized  form of the Black-Scholes partial differential equation, 
a closed form option pricing formula, and many other  results of 
mathematical finance  can be derived much as
is done in the standard theory.

Results generated for the behaviour of the price of a European call option
seem to capture some well-known features of real option prices. For example,
relative to the standard Black-Scholes model we find that a $q = 1.5 $ model
gives  a higher  value to  both  in-the-money and out-of-the-money options. 
This means
that option prices are quite well modelled using $q=1.5$ and  just one 
 value of the volatility parameter $\sigma$ across all strikes.  
As a result of this, we find that  the implied volatilities needed for a standard Black-Scholes model ($q=1$) to match the $q=1.5$ model show  a smile feature across strikes 
which  qualitatively behaves  much like empirical observations. 

To get a feel for the Greeks of our model, the dependency of the call price on 
each variable was calculated and plotted 
for values $q=1$ and $q=1.5$. In addition, we introduce a new Greek 
$\Upsilon$ to represent the variation of the option price with respect to
the parameter $q$. Option prices and partials do deviate significantly
 from the standard $q=1$ case as $q$ increases. 

Furthermore, we have  implemented numerical pricing routines which can be used
both for European and
 American options.  These entail implicitly solving the generalized Black Scholes 
differential equation Eq (\ref{eq:bsgen}). Results  from both methods agree very well, and were
further confirmed  by calculations involving monte carlo simulations of the
underlying stochastic process for the returns. 

We must yet study whether the prices obtained for American options 
match  observed ones.  However, based on the initial results obtained for European 
options we  are hopeful that this will be the case. We found that $q=1.4$ with 
one value of $\sigma$ across all strikes matches market prices extremely well, at least for
 the case of
calls on Japanese Yen futures studied here, with times to expiration ranging from 17 to 
147 days. In this example, the volatility surface 
for a standard Black
model ($q=1$)  is curved  across strikes with a slightly  upward trending term structure,
 while the one found 
for $q=1.4$ is flat across strikes with  a slightly downward trending term structure. 
Empirical work  is still  required 
to see if better option
replication can be achieved, and if arbitrage opportunities can be uncovered that do not
appear when the standard model is used. The pricing of exotic options is another 
topic open for future study.

In closing, we'd like to point out that this work is a first attempt at developing 
a theory of option pricing based on
a noise process evolving according to a nonlinear Fokker-Planck equation.
We have assumed that the parameter $q$ is constant for the evolution of 
returns across  all timescales,
but one natural theoretical extension of this work could be to let $q$ be a 
function of the timescale. Another possible extension to the model would be to include an 
assymmetry in the underlying noise distribution.

\appendix
\section {Appendix}

A consequence of Ito stochastic calculus is  the noise induced drift which
appears whenvever a  transformation of variables occurs.  
How does this noise induced drift  look, and what are its implications,
 in the generalized case? In the standard case, there are two common and equivalent 
 starting
points for modelling the dynamics of stock returns.  In the current framework
both of those starting points are also valid, and give identical 
option pricing results. Here we briefly depict the generalized versions
of these two models :  

One possible starting point start is  as in Eq(\ref{eq:langenlnS}), namely   
\begin{equation}
%\label{eq:langenlnS}
d Y = \mu dt  + \sigma  d \Omega
\end{equation}
with $Y(t) = \ln S(t+ \tau)/S(\tau)$ and  $d \Omega$ given by Eq(\ref{eq:nomega})
yielding 
\begin{eqnarray}
%\label{eq:langen1}
d S& = &(\mu + \frac{\sigma^2}{2}P_q^{1-q}) Sdt  + \sigma S d \Omega\\
& = &\tilde{\mu} Sdt  + \sigma S d \Omega
\end{eqnarray}
Here, the $\sigma^2/2 P^{1-q}_q$ term is a consequence if Ito's Lemma, and
corresponds to the noise induce drift term.
Alternatively we could choose to start with
\begin{equation}
\label{eq:langenS1}
d S = \mu  Sdt  + \sigma S d \Omega
\end{equation} 
as a model for the stock price evolution across the timescale $t$, resulting in 
\begin{equation}
\label{eq:langen2}
d Y = (\mu - \frac{\sigma^2}{2} P_q^{1-q}) dt  + \sigma  d \Omega
\end{equation} 
With Eq(\ref{eq:langenS1}) as a starting point, the noise induced drift term
enters in the equation  for $\ln S$.
The question now is deciding which model to use, Eq(\ref{eq:langenlnS}) 
or Eq(\ref{eq:langenS1})? We can derive option pricing formulas 
using either one as a starting point.   
 The good news is that, just as in the 
standard case (recovered for $q=1$),  the two models yield identical 
option pricing formulas, because either way, $\mu$ or $\tilde{\mu}$  disappears under the equivalent martingale measure.

\section {Appendix}

The formula for $S(T)$ (Eq(\ref{eq:SF}) or Eq(\ref{eq:mstock}) contains terms of type
\begin{equation}
\int_0^T P(\Omega(s),s) ^{1-q} ds
\end{equation}
But for each value of  $s$, the distribution of the random variable
$\Omega(s)$ follows a Tsallis distribution of the form
\begin{equation}
P_q(\Omega(s),s) = \frac{1}{Z(s)} ( 1 - (1-q) \beta(s)\Omega(s)^2)^{\frac{1}{1-q}}
\end{equation}
Each such distribution can be mapped onto the distribution of a standardized
random variable $x_N$ through the variable transformation
\begin{equation}
\label{eq:trafo1}
x_N = \sqrt{\frac{\beta(s)}{\beta_N} } \Omega(s)
\end{equation}
with distribution
\begin{equation}
P_q(x_N) = \frac{1}{Z_N} ( 1 - (1-q) \beta_N x_N^2)^{\frac{1}{1-q}}
\end{equation}
where the standard relation
\begin{equation}
P_q(x_N) = P_q(\Omega(s),s) \frac{\partial \Omega_s}{\partial x_N}
\end{equation}
holds.
Note that we can in turn map the standardized distribution of $x_N$  
onto the distribution of the variable $x(T)$ at the fixed timescale $T$ via
the variable transformation 
\begin{equation}
\Omega(T) = \sqrt{\frac{\beta_N}{\beta(T)} } x_N
\end{equation}
This result could have also  been achieved directly via the
 variable transformation
\begin{equation}
\Omega(s)  = \sqrt{\frac{\beta(T)}{\beta(s)}} \Omega(T)
\end{equation}

\bigskip

\begin{figure}[t]
\psfig{file=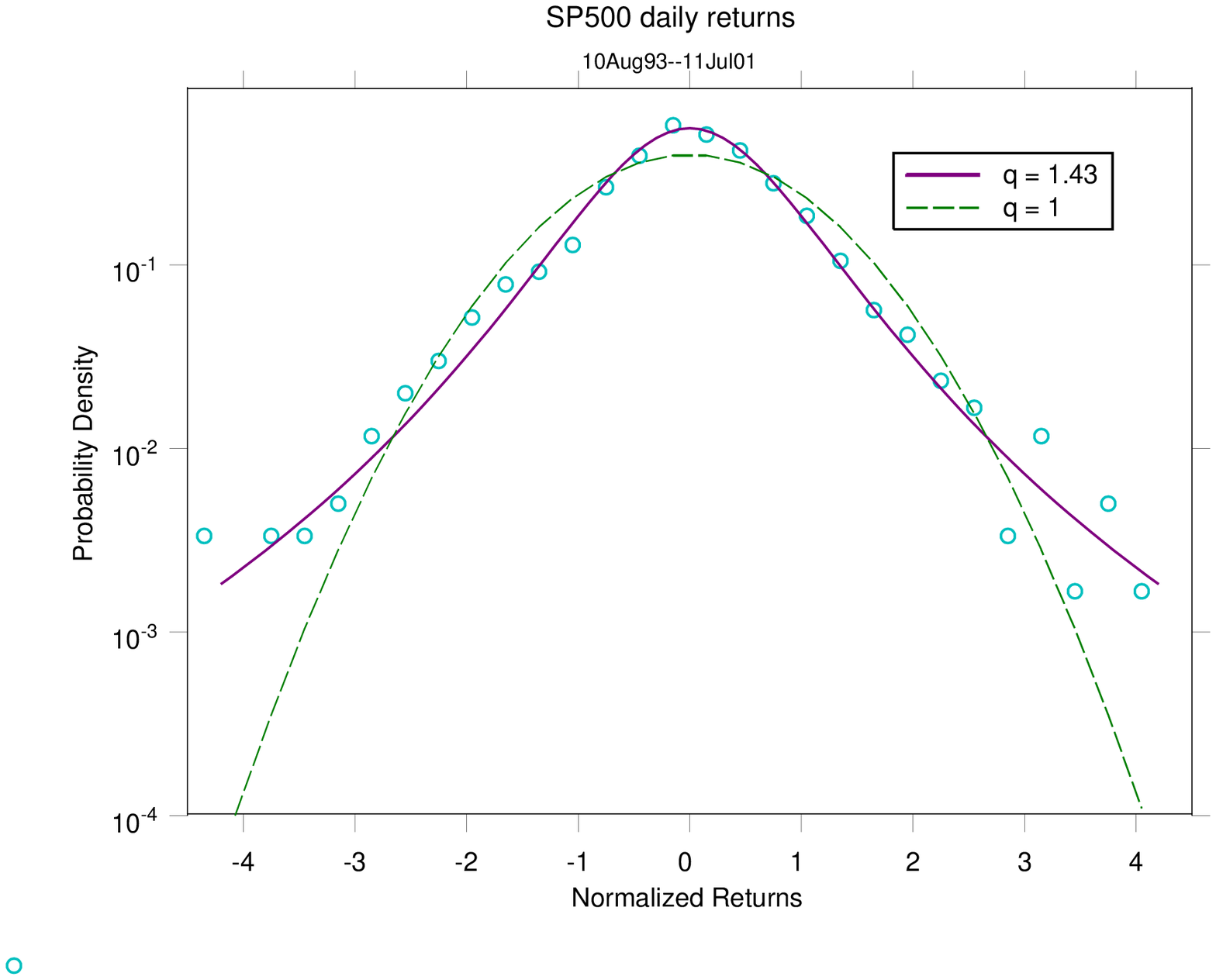,width=4.5in}
\caption{\footnotesize
Distributions of log returns, normalized by the sample standard deviation,  rising from the demeaned S \& P 500,
and from  a Tsallis distribution of index $q = 1.43$ (solid line). For comparison, the  normal distribution
is also shown ($q= 1$, dashed line). Figure kindly provided by R. Osorio, to be published 
\cite{osorio}}
\end{figure}

\begin{figure}[t]
\psfig{file=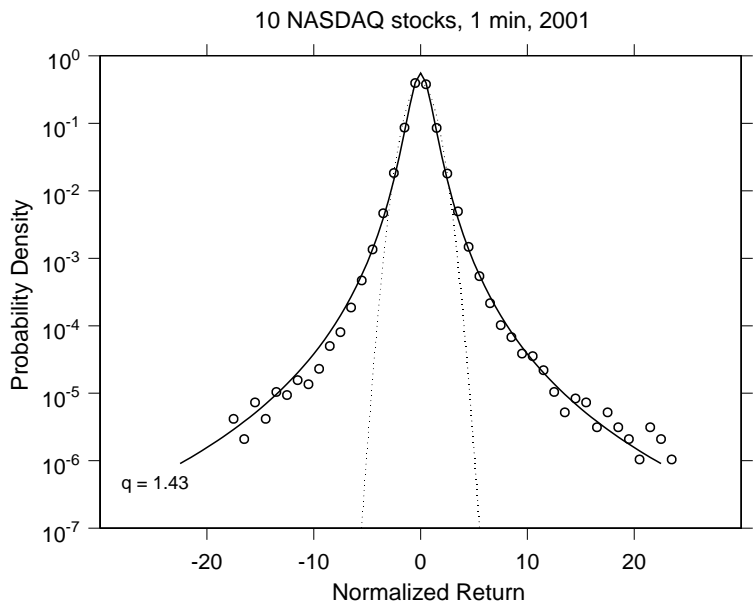,width=4.5in}
\caption{\footnotesize
Distributions of log returns for  10 Nasdaq  high-volume
 stocks. Returns are calculated over 1 minute intervals, and are normalized by the sample standard deviation. Also shown is the Tsallis distribution of index $q = 1.43$ (solid line) which provides a good fit to the data. 
Figure kindly provided by R. Osorio, to be published \cite{osorio}}
\end{figure}

\bigskip

\begin{figure}[t]
\psfig{file=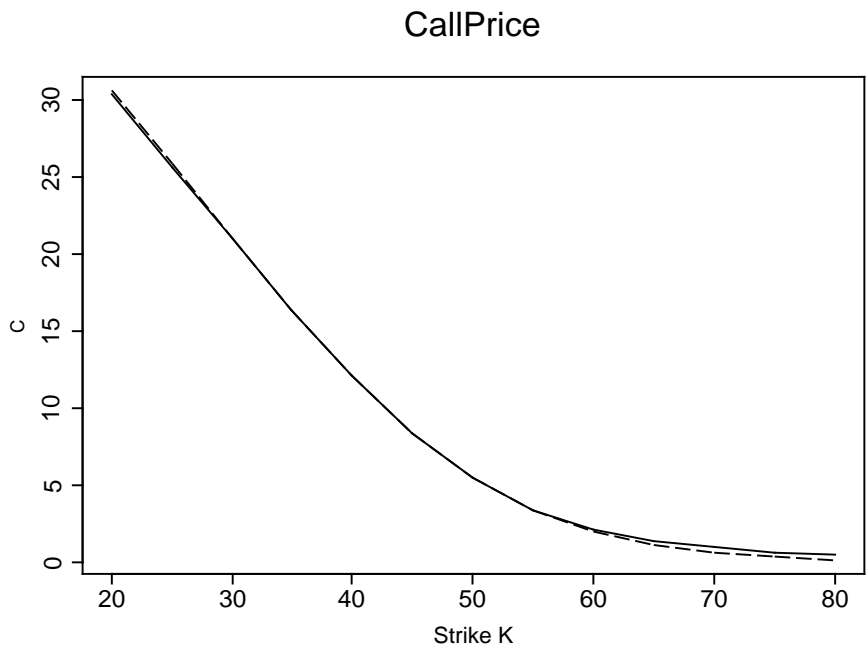,width=4.5in}
\caption{\footnotesize
Call option price versus strike price, using $S(0) = 50$, $r = .06$ and $T=.6$,
for $q=1$ (dashed curve) and $q=1.5$ (solid curve).
For each  $q$, $\sigma$ was chosen so that the at-the-money options 
are priced equally ($\sigma =.3$ for $q=1$  and $\sigma=.299$ for $q=1.5$).  
}
\end{figure}

\bigskip

\begin{figure}[t]
\psfig{file=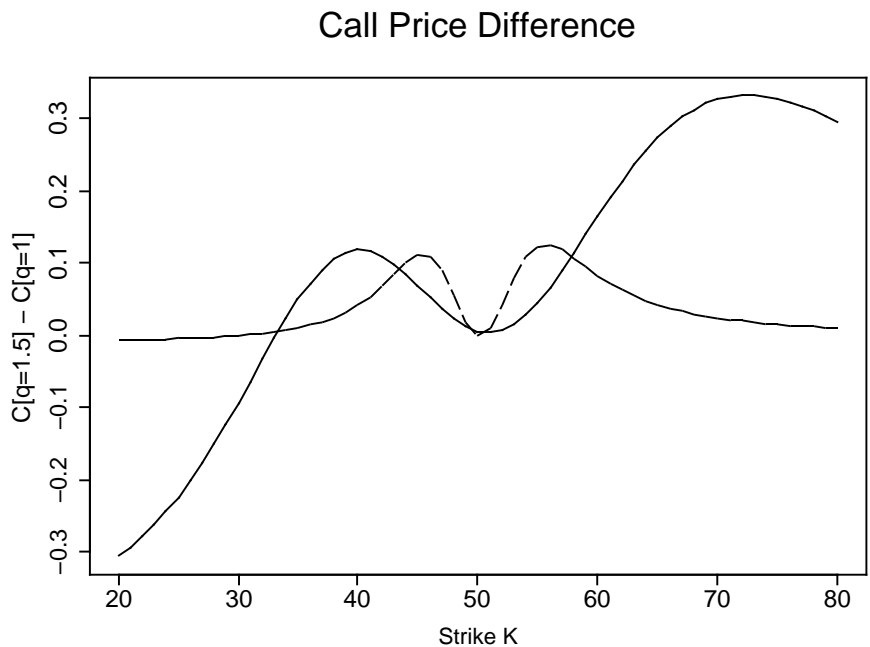,width=4.5in}
\caption{\footnotesize
Calibrated so that at-the-money options are priced equally, the difference
between the $q=1.5$ model and the standard Black-Scholes model is shown, for  $S(0) = 50\$$ and  $r= 0.06$. The solid line corresponds to $T=0.6$  with $\sigma = .3$
for $q=1$ and  $\sigma = .297$  for $q=1.5$. The dashed line represents $T=0.05$ 
with $\sigma = .3$ for $q=1$ and $\sigma = .41$ for  $q=1.5$.  Times are 
expressed in years, $r$ and $\sigma$ are in annual  units.}
\end{figure}

\bigskip

\begin{figure}[t]
\psfig{file=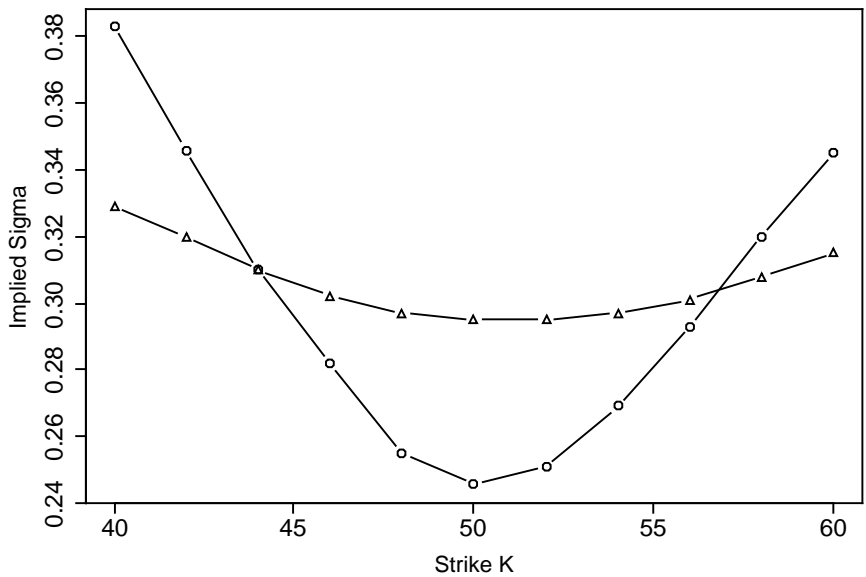,width=4.5in}
\caption{\footnotesize
Using the $q=1.5$ model (here with $\sigma =.3$, $S(0) = 50 \$$ and $r=.06$) to generate call option prices, one can back out
the volatilities implied by a standard $q=1$ Black-Scholes model.
Circles correspond to $T=0.4$, while
triangles represent $T= 0.1$.
These implied volatilities capture
features seen in real options data. In particular, the smile is more 
pronounced for small $T$. 
}
\end{figure}

\bigskip

\begin{figure}[t]
\psfig{file=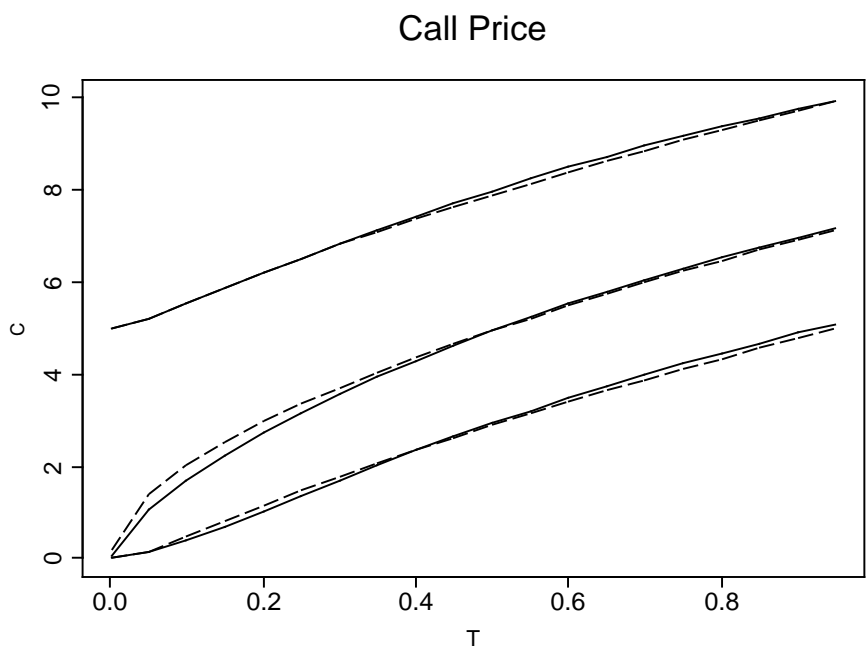,width=4.5in}
\caption{\footnotesize
Call option price versus time to expiration, using $S(0) = 50$, $r = .06$, and $\sigma = .3$. Three different strikes were considered: $K = 45$ (in-the-money, top), $K = 50$ (at-the-money, middle) and $K=55$ (out-of-the-money, bottom).
 Curves for $q= 1$ (dashed) and $q=1.5$ (solid) are shown.  
}
\end{figure}

\bigskip

\begin{figure}[t]
\psfig{file=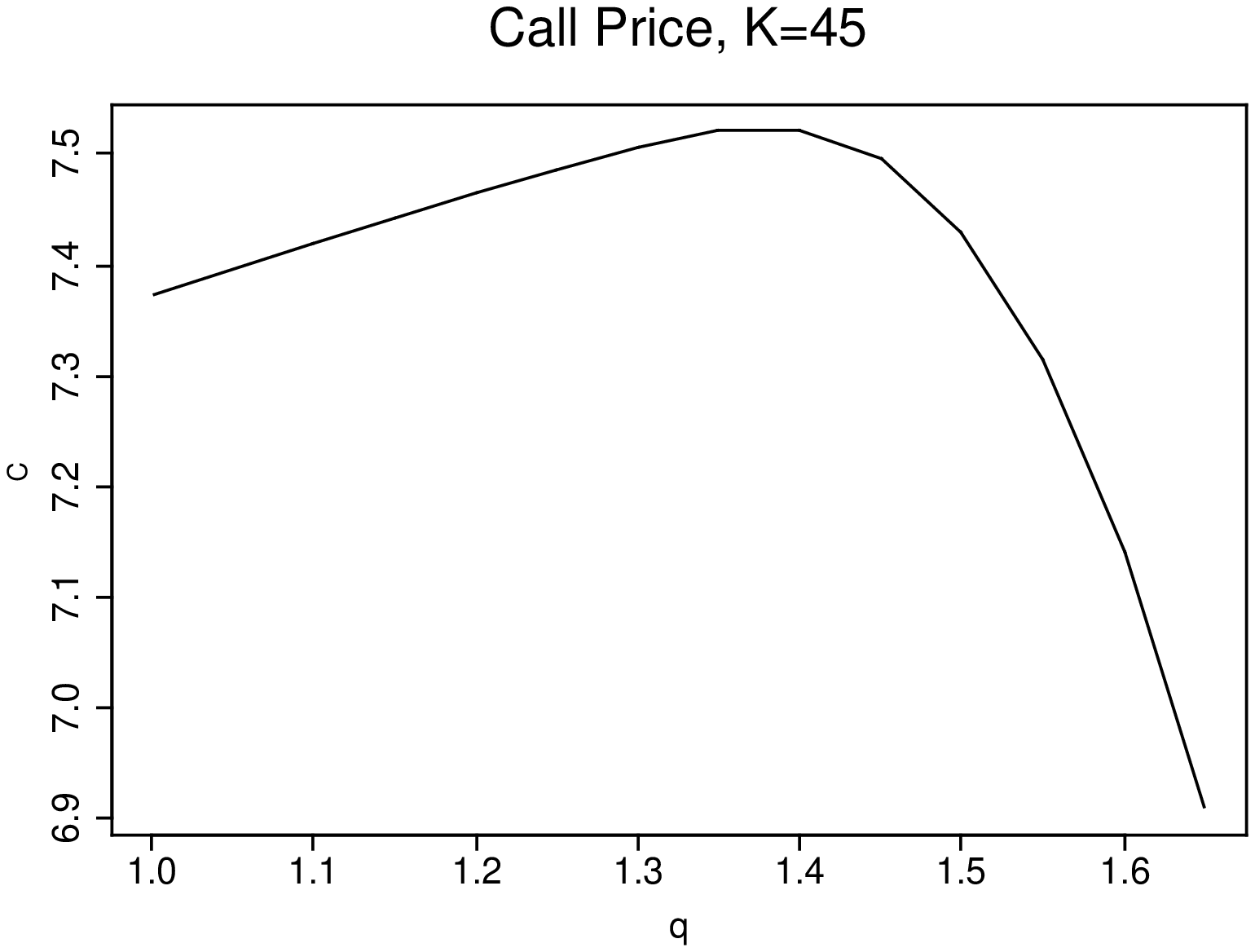,width=2.5in}
\psfig{file=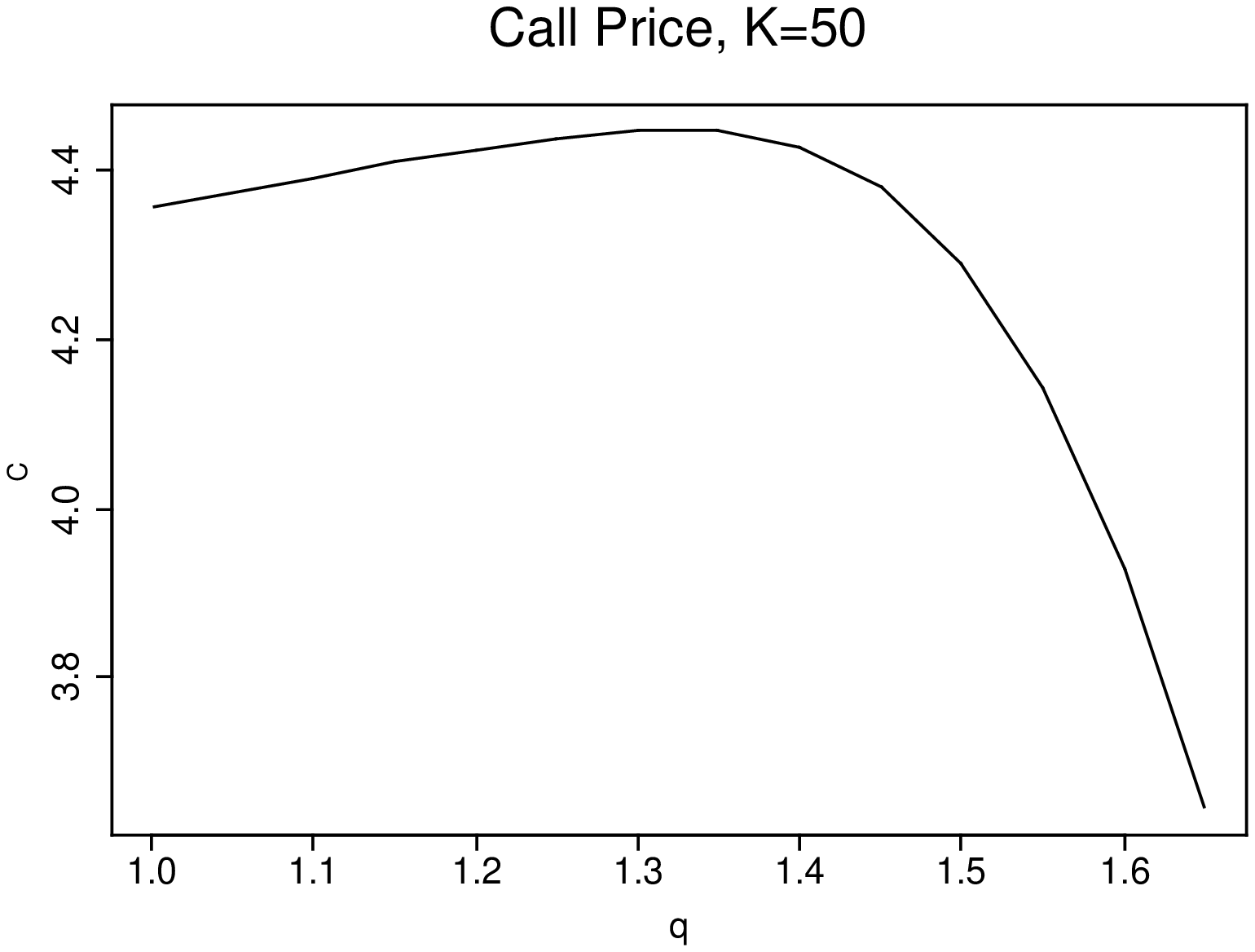,width=2.5in}
\psfig{file=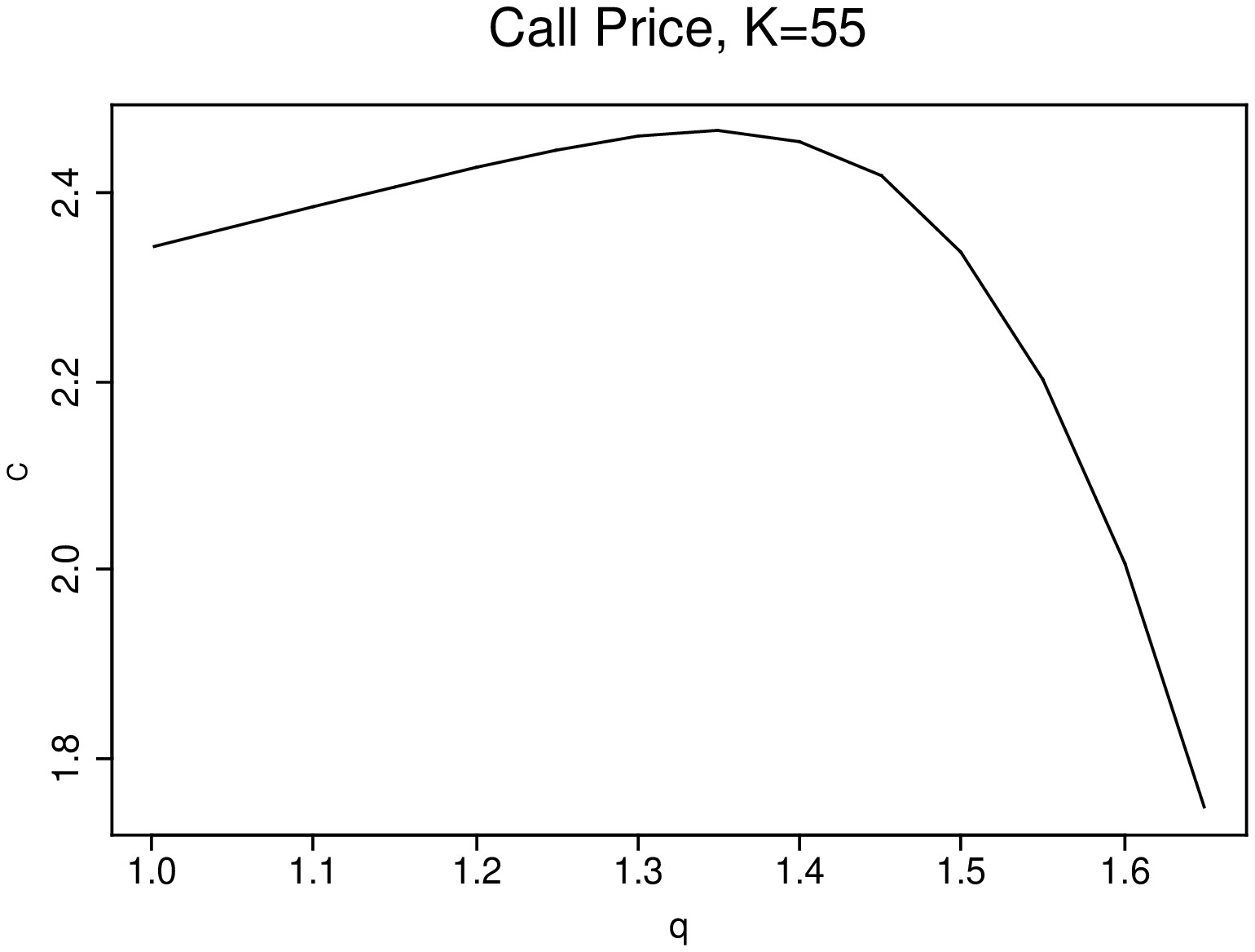,width=2.5in}
\caption{\footnotesize
Call option price versus $q$, using $S(0) = 50$, $r = .06$, and $T=0.4$. 
Three different strikes were considered: $K = 45$ (in-the-money, top), $K = 50$ (at-the-money, middle) and $K=55$ (out-of-the-money, bottom).   
}
\end{figure}

\bigskip
\begin{figure}[t]
\psfig{file=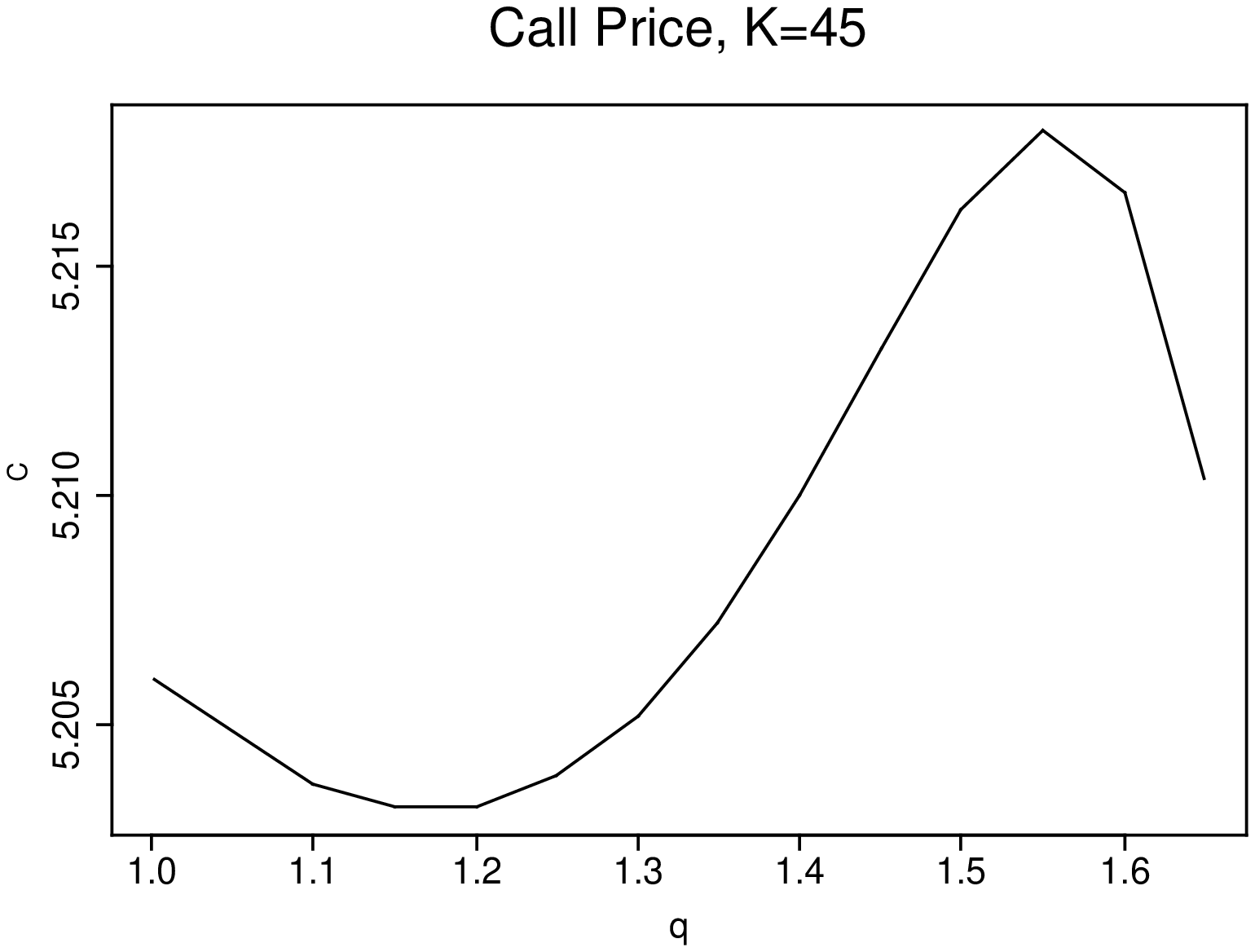,width=2.5in}
\psfig{file=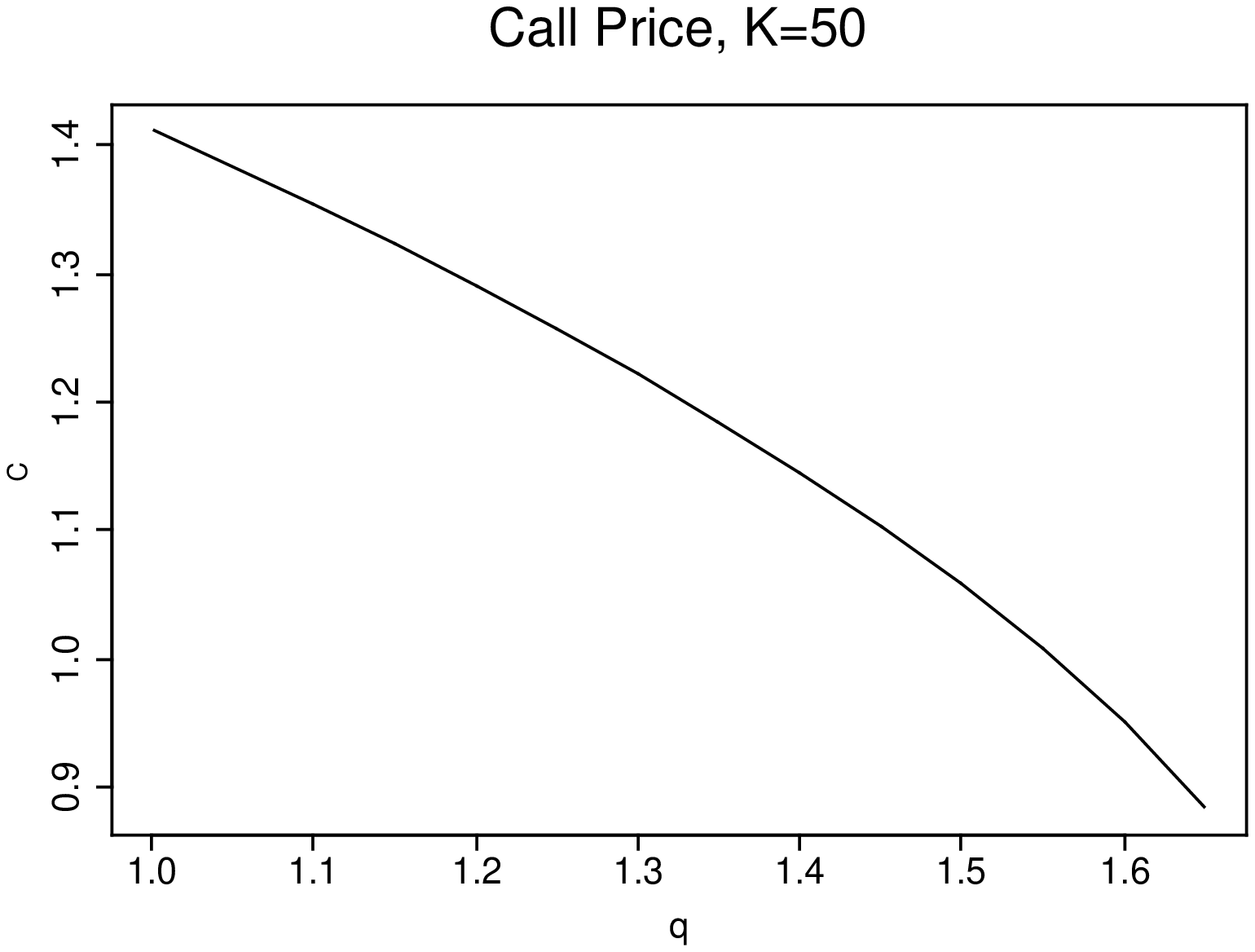,width=2.5in}
\psfig{file=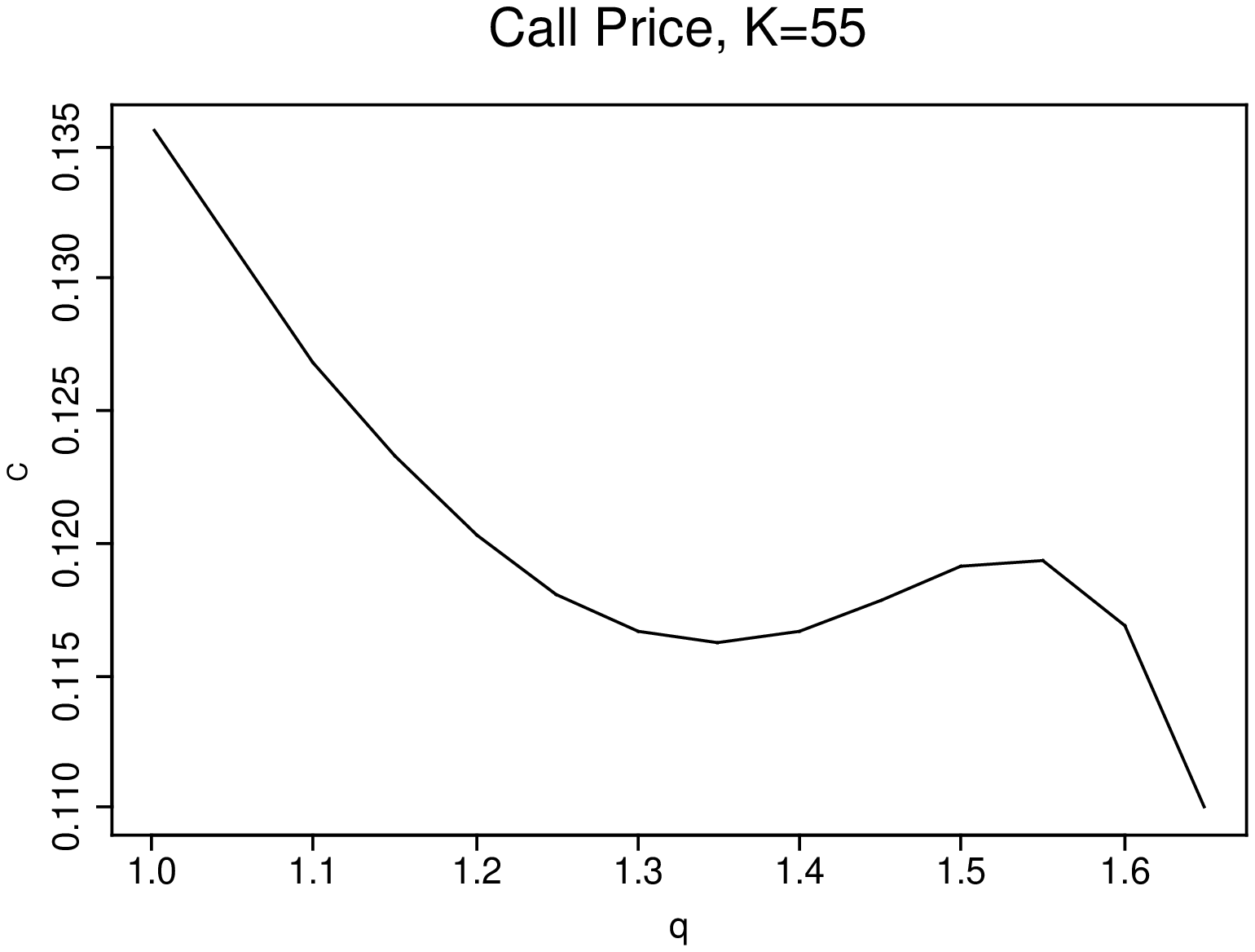,width=2.5in}
\caption{\footnotesize
Call option price versus $q$, using $S(0) = 50$, $r = .06$, and $T=0.05$. 
Three different strikes were considered: $K = 45$ (in-the-money, top) $K = 50$ (at-the-money, middle) and $K=55$ (out-of-the-money, bottom).   
}
\end{figure}

\bigskip

\begin{figure}[t]
\psfig{file=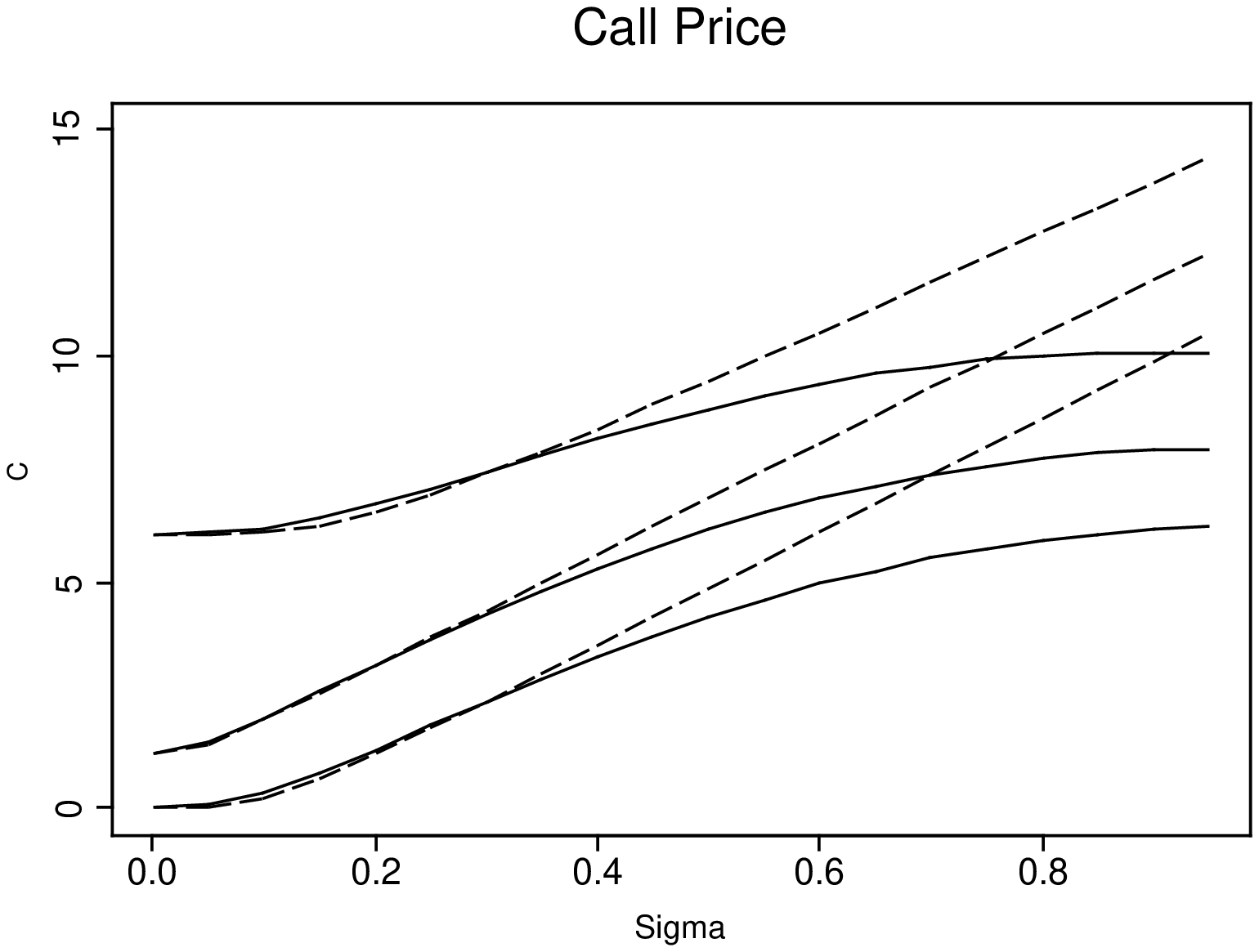,width=4.5in}
\caption{\footnotesize
Call option price versus $\sigma$, using $S(0) = 50$, $r = .06$, and $T=0.4$. 
Three different strikes were considered: $K = 45$ (in-the-money, top), $K = 50$ (at-the-money, middle) and $K=55$ (out-of-the-money, bottom). 
Curves for $q= 1$ (dashed) and $q=1.5$ (solid) are shown.  
}
\end{figure}

\bigskip

\begin{figure}[t]
\psfig{file=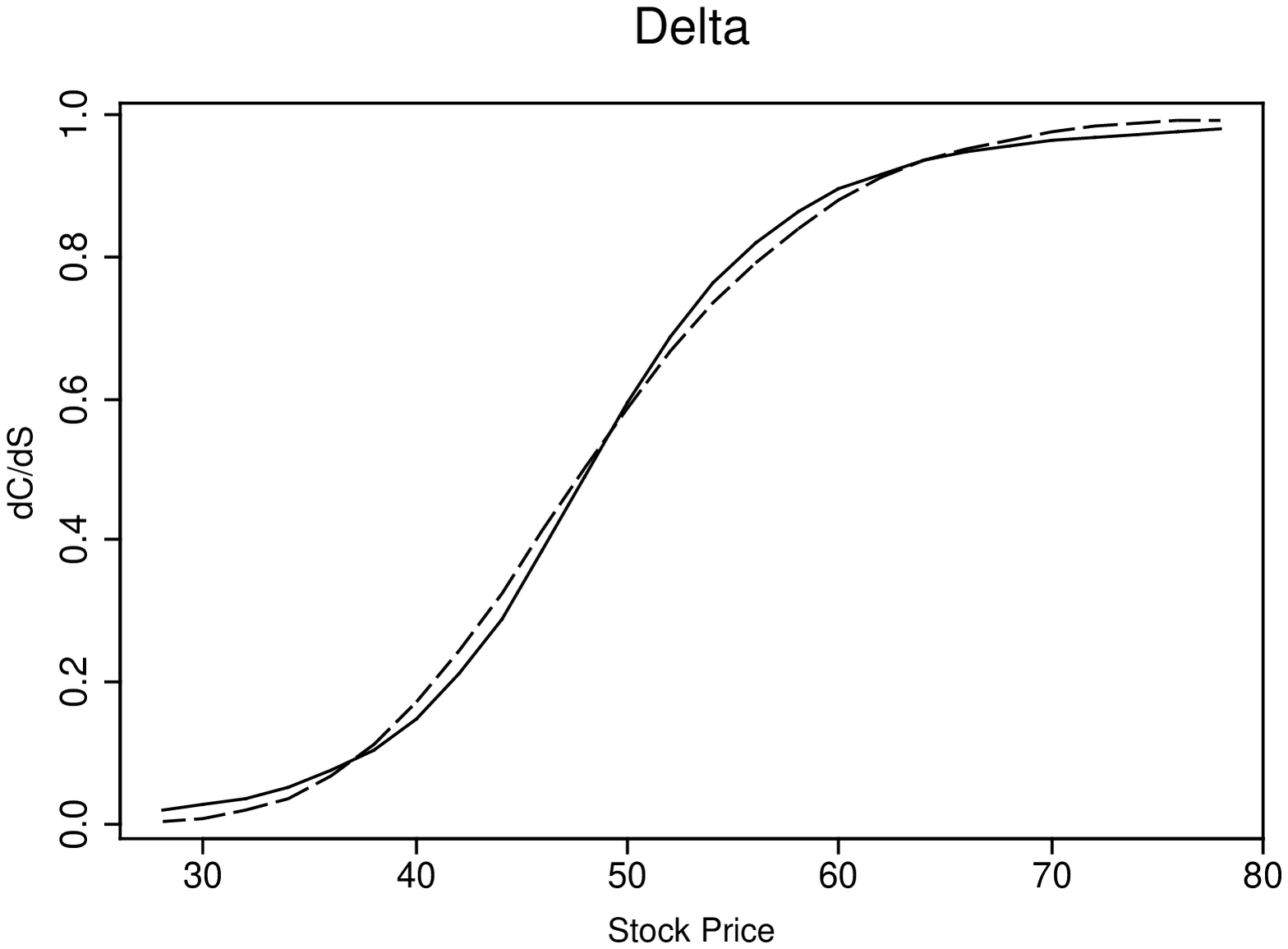,width=4.5in}
\caption{\footnotesize
 $\Delta = \frac{\partial c}{\partial S}$ as a function of the stock price 
$S = S(0)$  using $K = 50$, $r = .06$, and $T=0.4$. Curves for $q =1 $ (dashed) and $q=1.5$ (solid)  are shown. 
}
\end{figure}

\bigskip

\begin{figure}[t]
\psfig{file=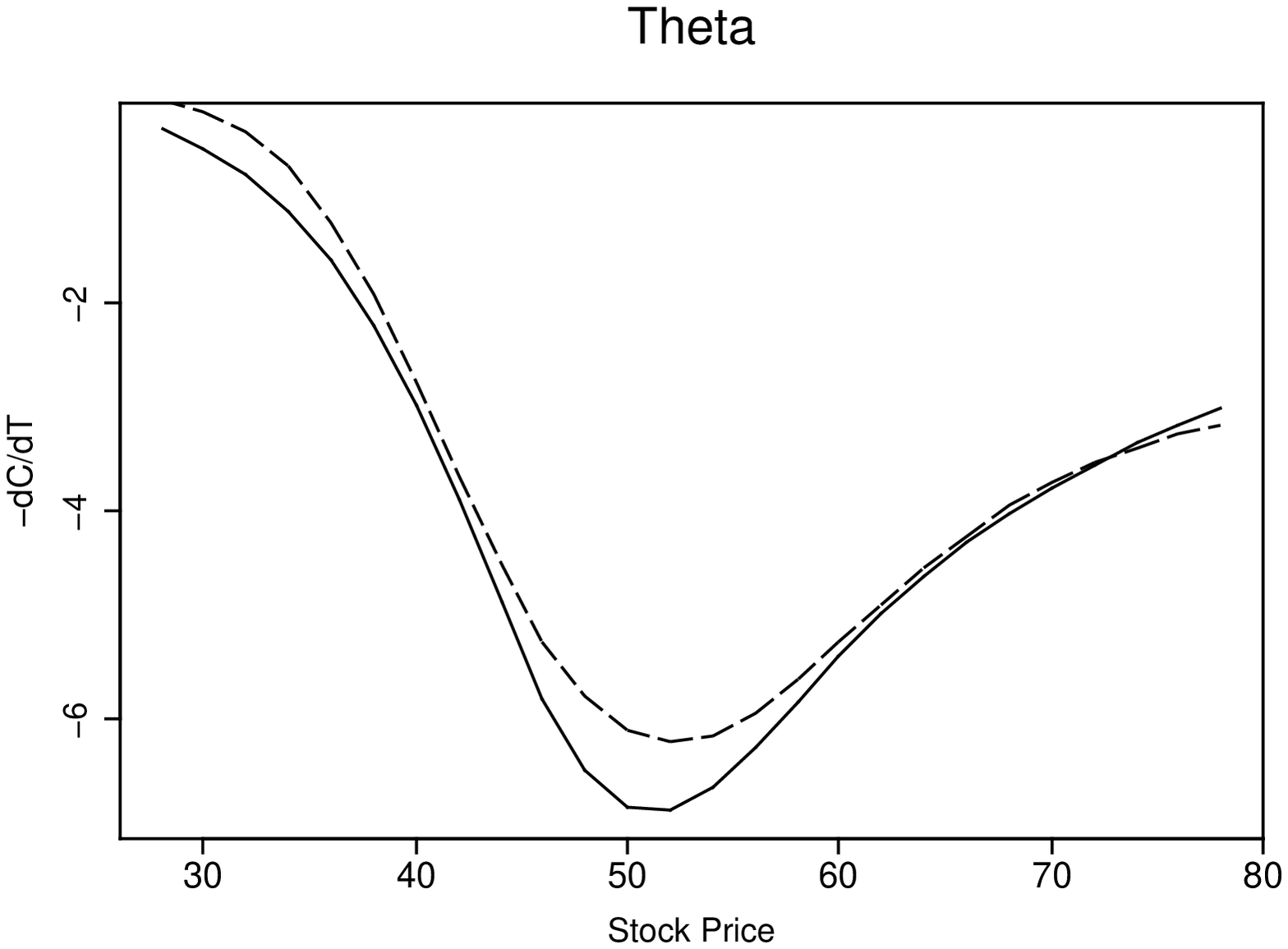,width=4.5in}
\caption{\footnotesize
 $\Theta = -\frac{\partial c}{\partial T}$ as a function of the stock price 
$S = S(0)$  using $K = 50$, $r = .06$, and $T=0.4$. Curves for $q =1 $ (dashed)  
 and $q=1.5$  (solid)  are shown. 
}
\end{figure}

\bigskip

\begin{figure}[t]
\psfig{file=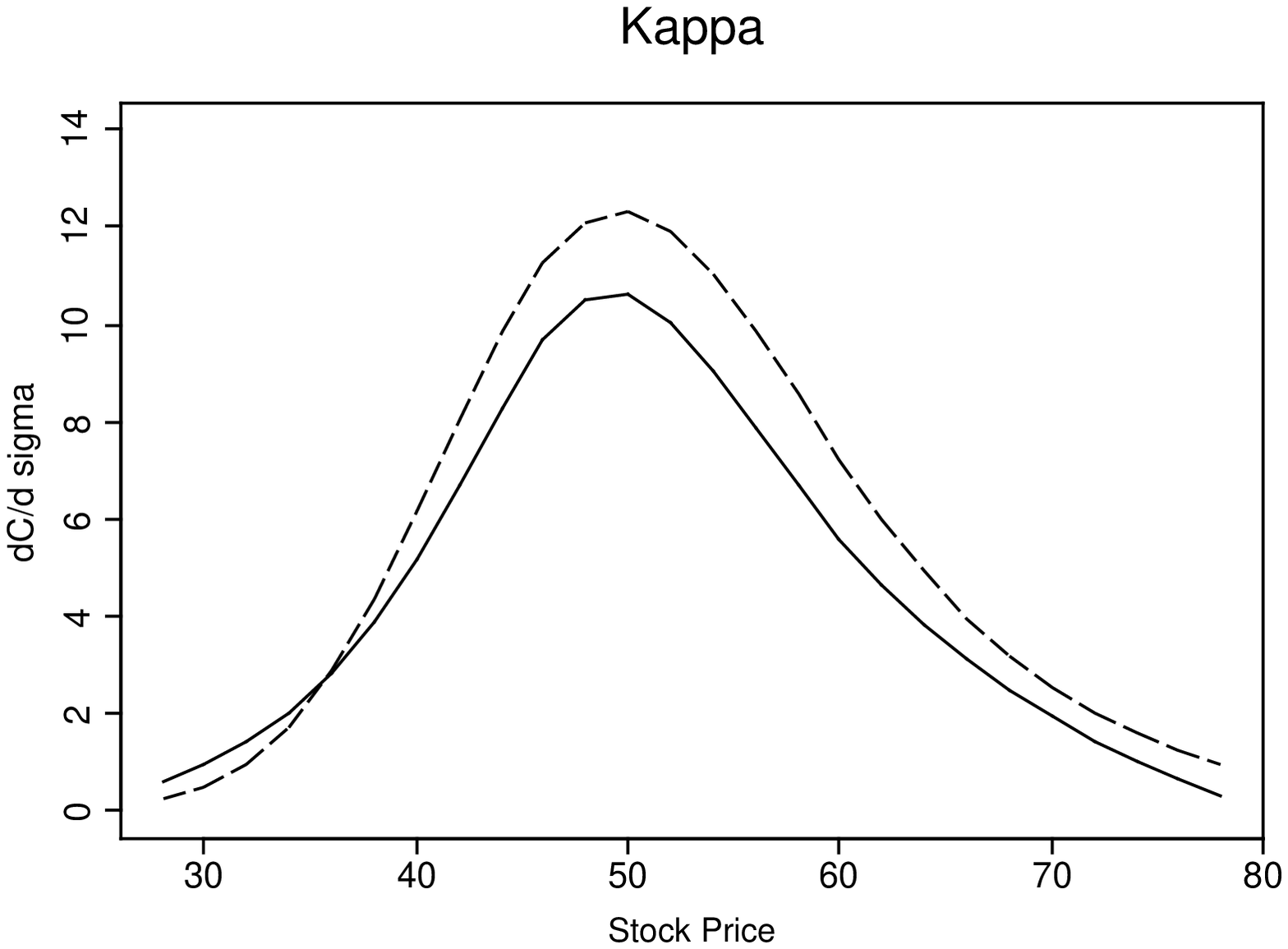,width=4.5in}
\caption{\footnotesize
 $\kappa = \frac{\partial c}{\partial \sigma}$ as a function of the stock price 
$S = S(0)$  using $K = 50$, $r = .06$, and $T=0.4$. Curves for $q =1 $ (dashed) 
and $q=1.5$ (solid) are shown. 
}
\end{figure}

\bigskip

\begin{figure}[t]
\psfig{file=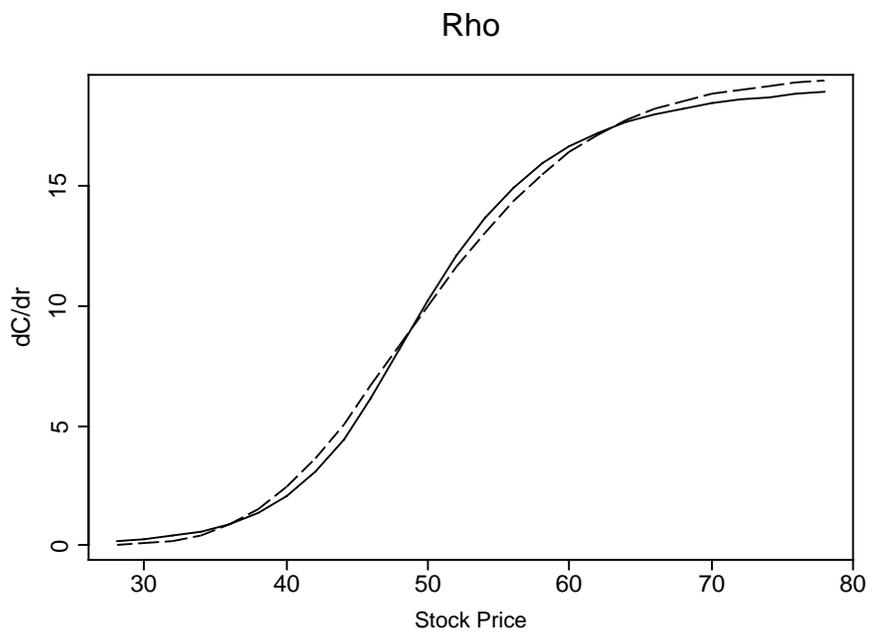,width=4.5in}
\caption{\footnotesize
 $\rho = \frac{\partial c}{\partial r}$ as a function of the stock price 
$S = S(0)$  using $K = 50$, $r = 0.06$, and $T=0.4$. Curves for $q =1 $ (dashed)  and $q=1.5$ (solid) are depicted. 
}
\end{figure}

\bigskip

\begin{figure}[t]
\psfig{file=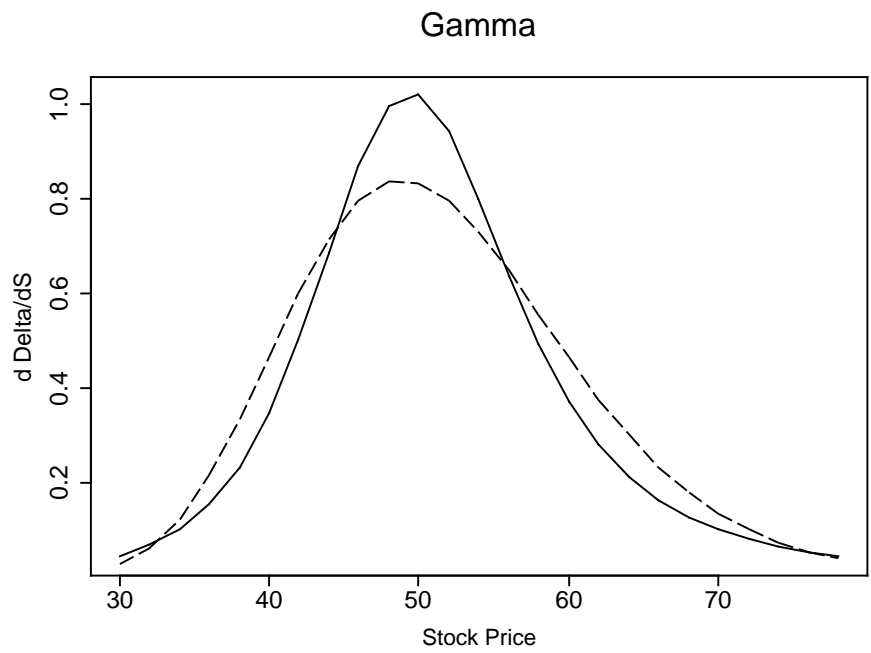,width=4.5in}
\caption{\footnotesize
 $\Gamma = \frac{\partial \Delta}{\partial S}$ as a function of the stock price 
$S = S(0)$  using $K = 50$, $r = 0.06$, and $T=0.4$. Curves for $q =1 $ (dashed)  and $q=1.5$ (solid) are depicted. 
}
\end{figure}

\bigskip

\begin{figure}[t]
\psfig{file=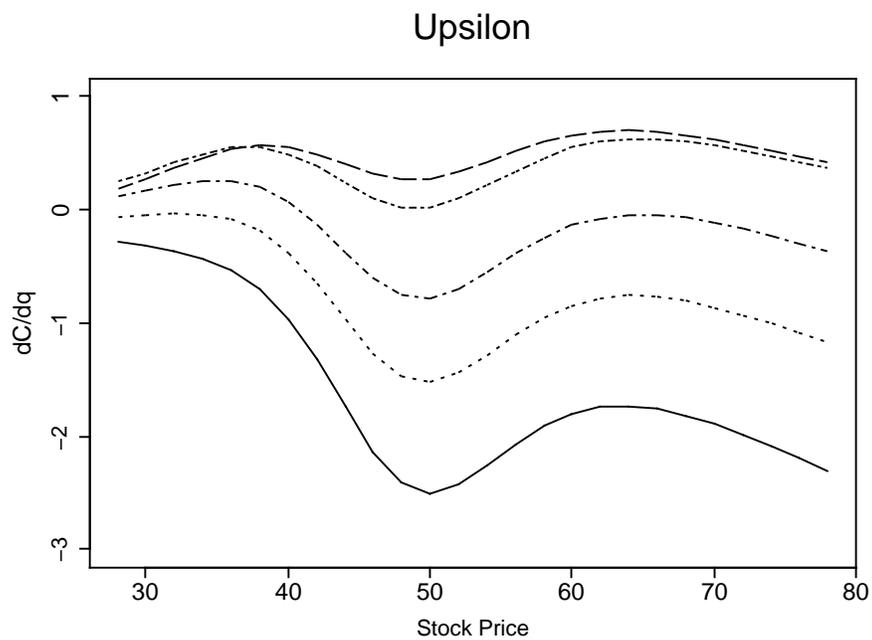,width=4.5in}
\caption{\footnotesize
 $\Upsilon = \frac{\partial c}{\partial q}$ as a function of the stock price 
$S = S(0)$  using $K = 50$, $r = .06$, and $T=0.4$. Curves for $q =1.1 $ (top,dashed) ranging to  $q=1.5$ (solid) are shown. The other curves correspond to $q=1.3$, $q=1.4$ and $q=1.45$ in order of descent. 
}
\end{figure}

\bigskip

\begin{figure}[t]
\psfig{file=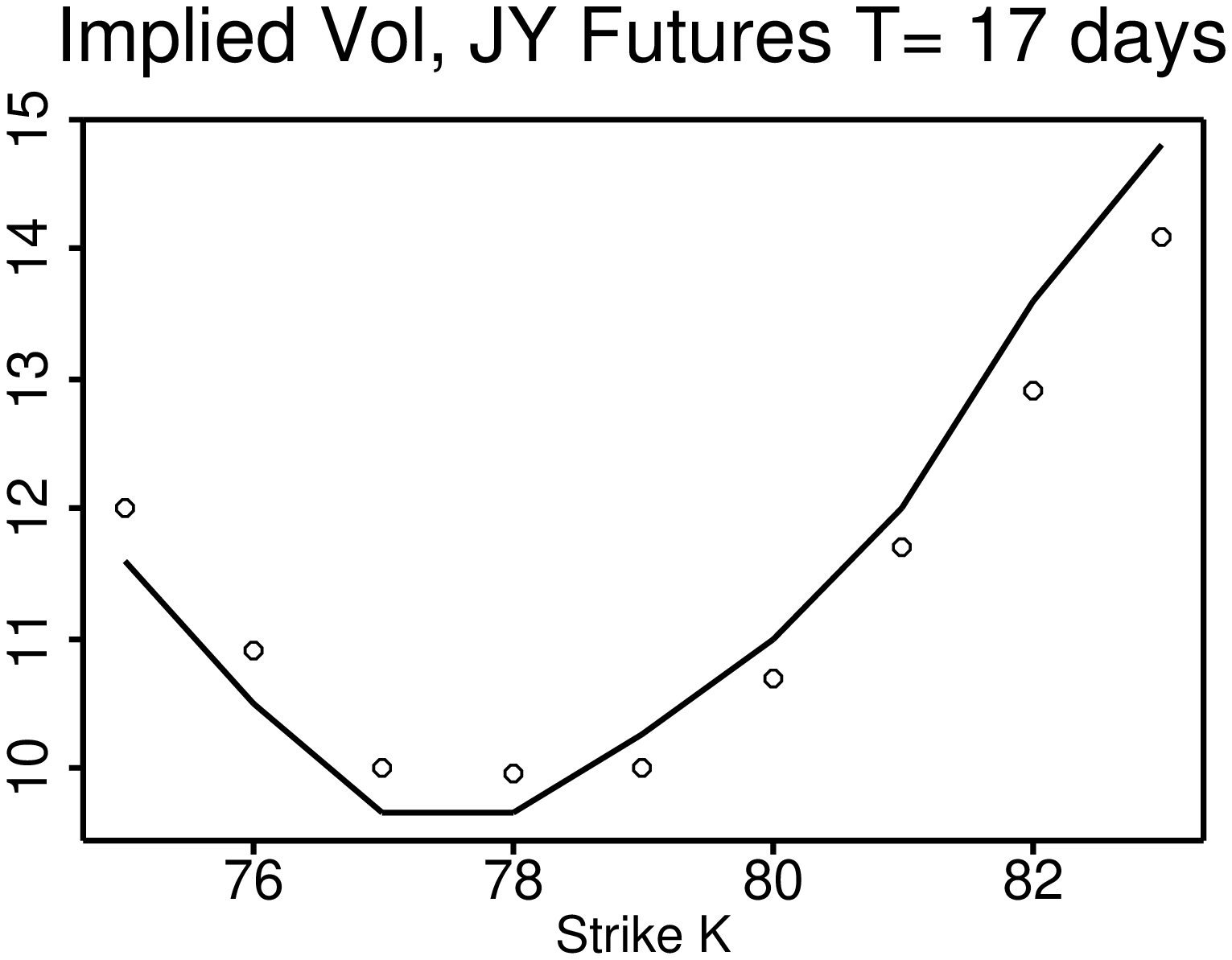,width=2.8in}
\psfig{file=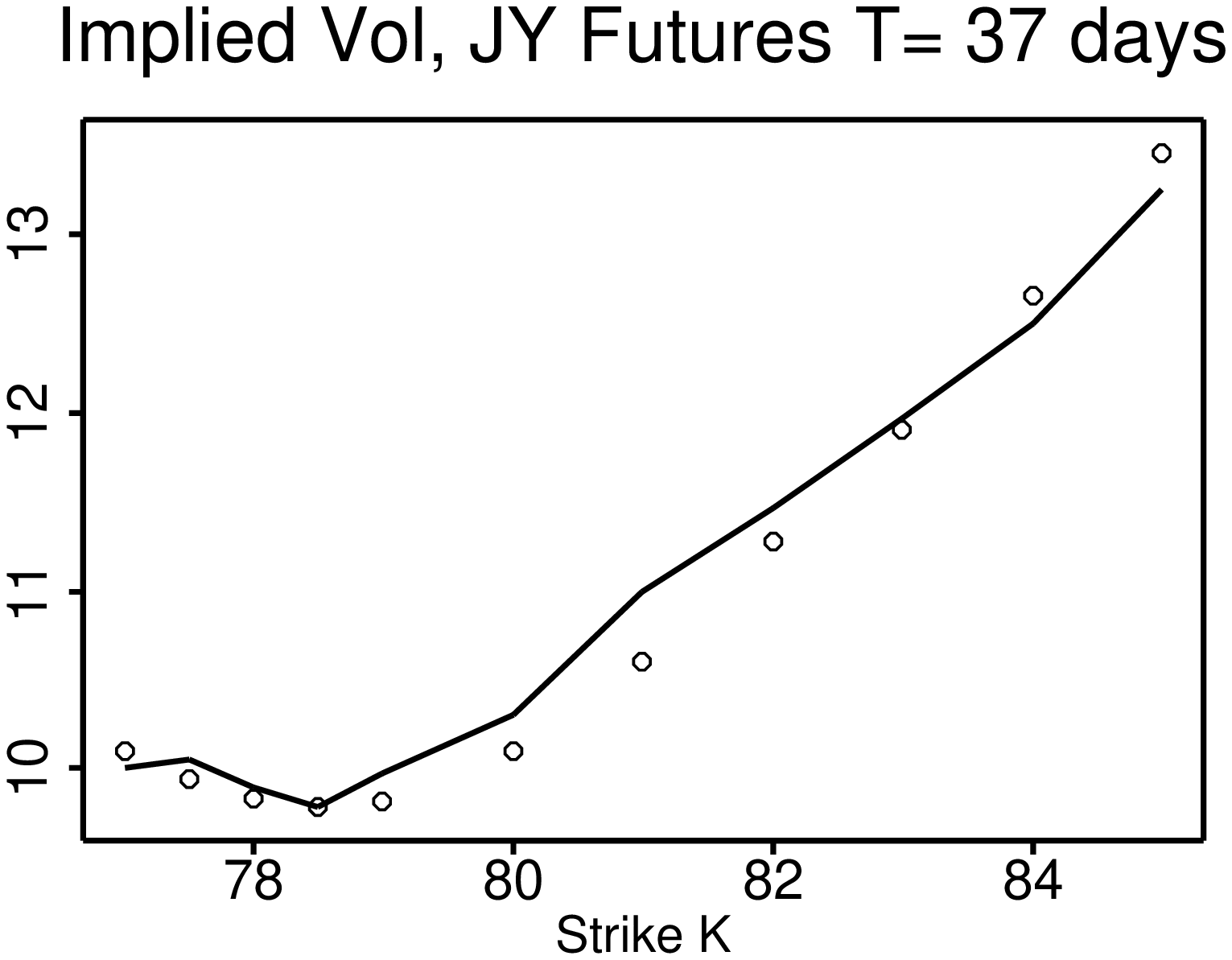,width=2.8in}
\caption{\footnotesize
Implied volatilies for call options on JY currency futures, traded on May 16 2002.
 The following plots will show how smiles implied from our model using $q=1.4$ match  market smiles for
 times to expiration ranging from about 2 weeks to close to half a year. 
Option prices were calculated using $q=1.4$ and just one value of $\sigma$ for each $T$ (chosen such that  the at-the-money option equals the market value).
The solid line corresponds to volatilities implied by the market. The symbols
correspond to volatilities implied by comparing a standard Black model to ours.
$ r =.055$ and 
Top)  $F(0) = 78.16$, $\sigma= 12.2\%$ and $T=17$ days.
Bottom)  $F(0) = 78.54$,$\sigma= 11.2\%$ and $T=37$ days.
}
\end{figure}

\bigskip

\begin{figure}[t]
\psfig{file=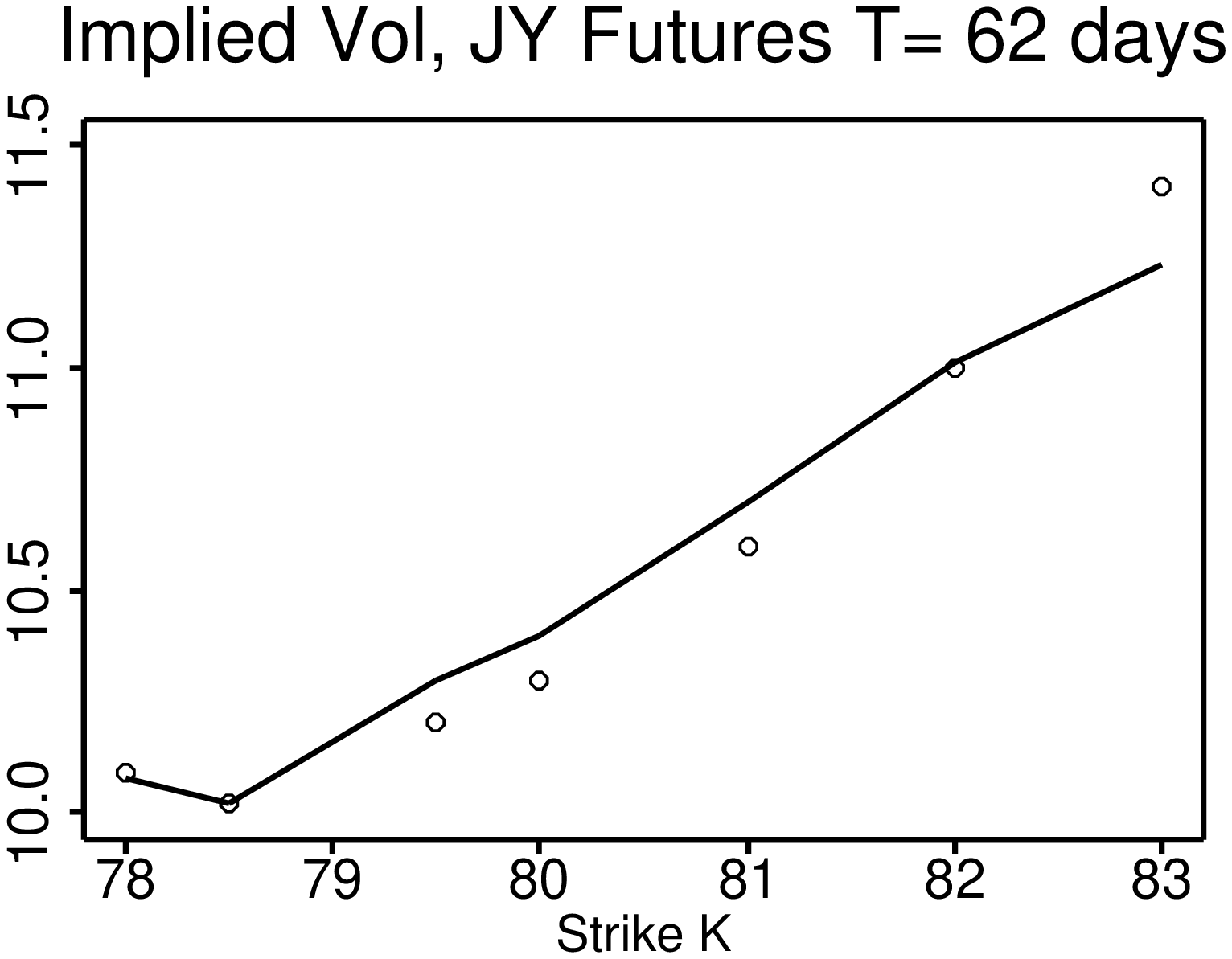,width= 2.8in}
\psfig{file=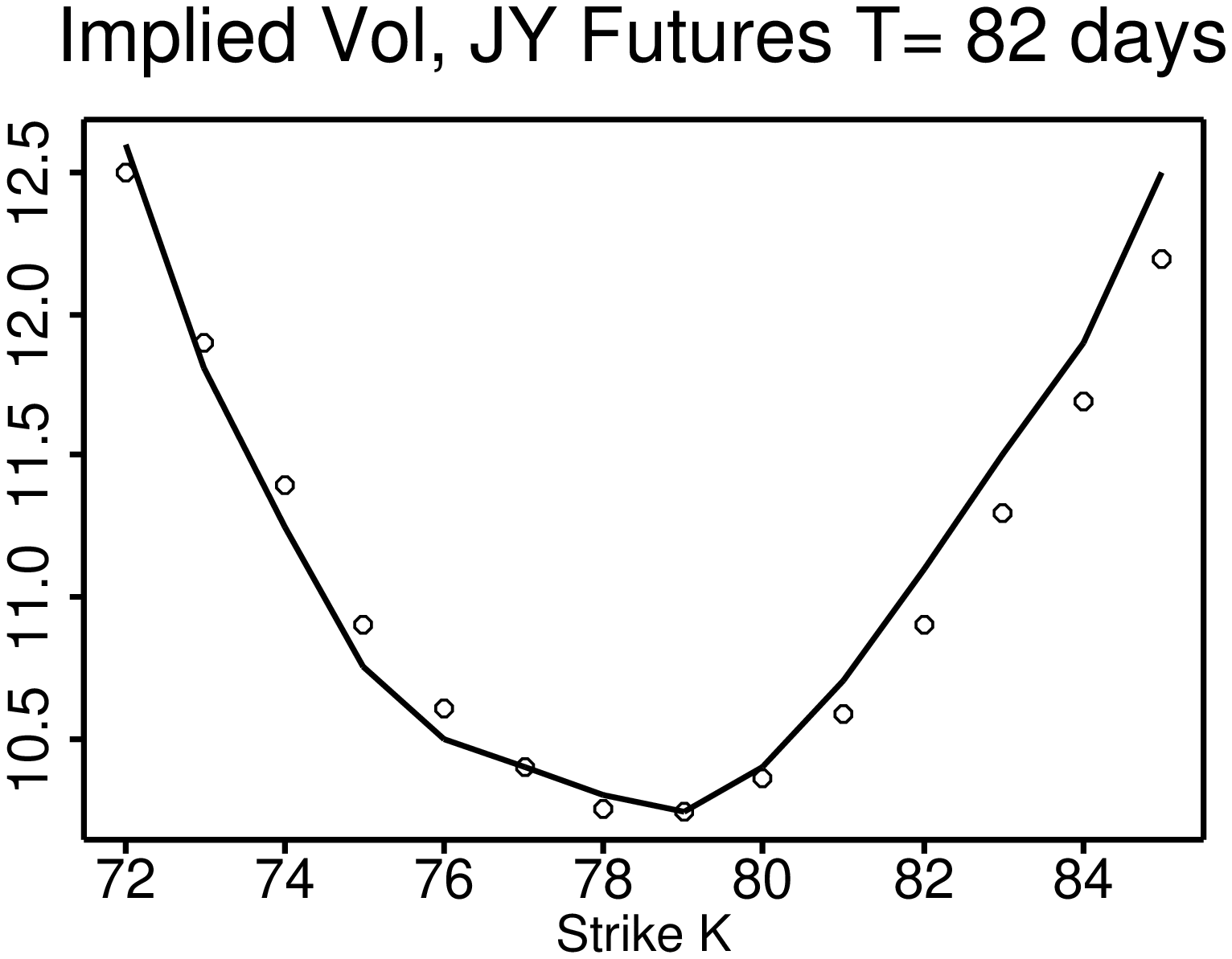,width= 2.8in}
\psfig{file=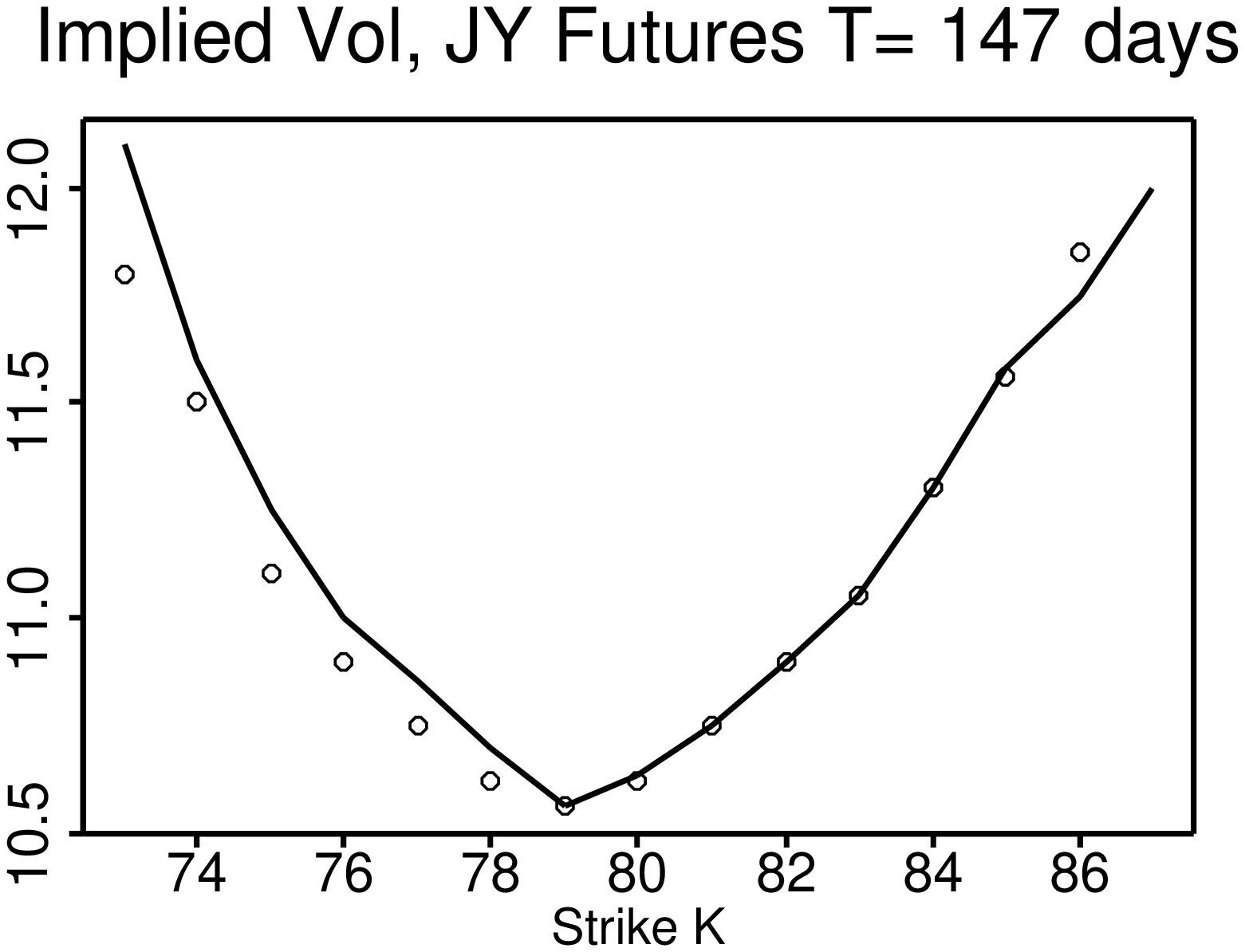,width=2.8in}
\caption{\footnotesize
Implied volatilies for call options on JY currency futures, traded on May 16 2002.
The solid line corresponds to market implied volatilities. Symbols
correspond to volatilities implied by our model with $q=1.4$,
$r=.055$ and
Top) $F(0) = 78.54$, $\sigma= 10.8\%$ and $T=62$ days.
Middle) $F(0) = 78.54$, $\sigma= 10.6\%$ and $T=82$ days.
Bottom) $F(0) = 79.01$,$\sigma= 10.2\%$ and $T=147$ days.
}
\end{figure}

\bigskip

\begin{figure}[t]
\psfig{file=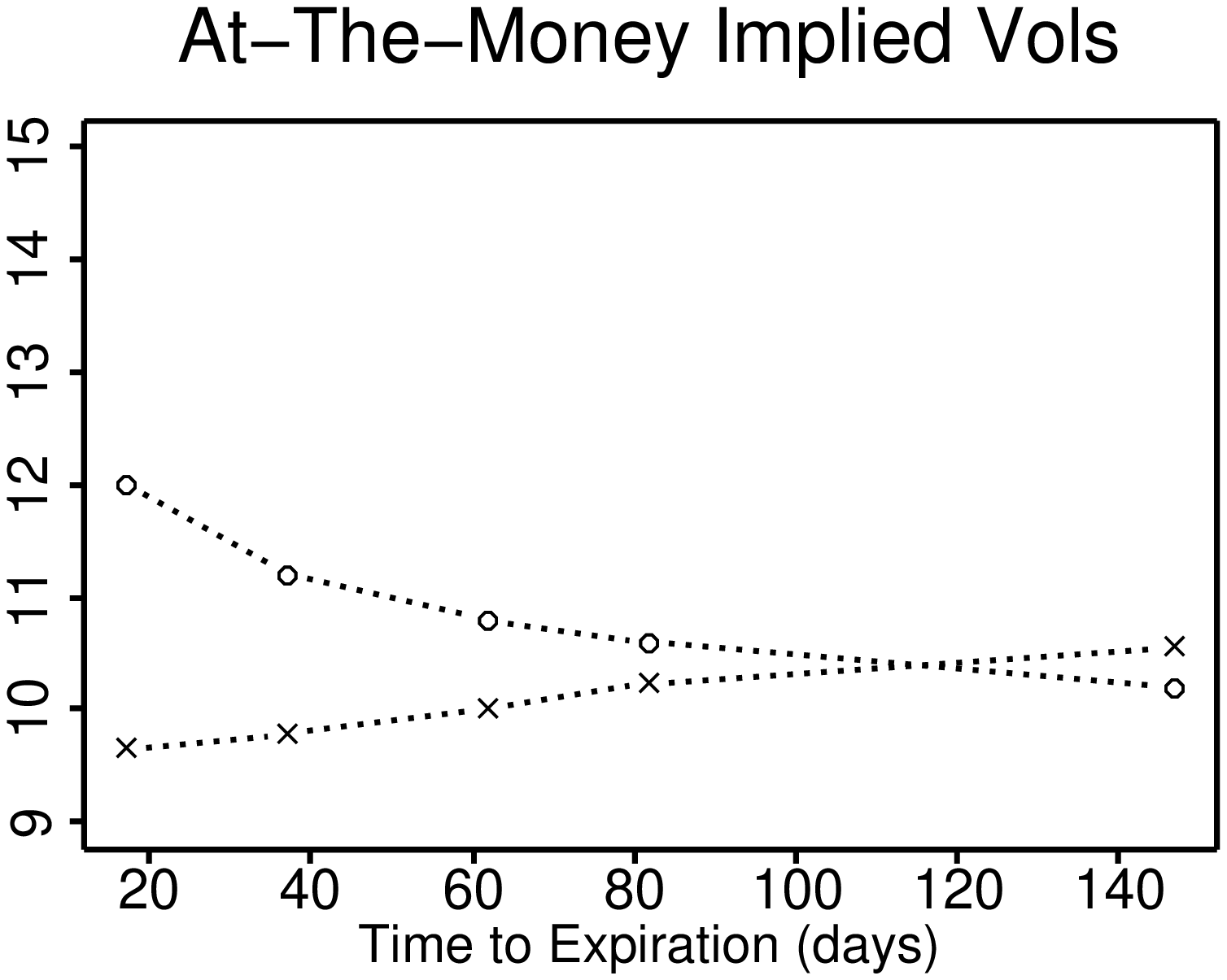,width= 4.5in}
\caption{\footnotesize
Term structure of the at-the-money volatility in the example of 
Figure 16 and Figure 17 above.  
The crosses correspond to at-the-money volatilities of a standard
Black model. Circles
correspond to at-the-money volatilities of the generalized  model with 
 $q=1.4$. While the volatility surface of the Black model consists of 
the evolution of smiles with the at-the-money volatility drifting  upwards with $T$,
  the volatility surface in the $q=1.4$ model is
simply given by a flat surface across strikes, drifting downwards with $T$.
}
\end{figure}

\bigskip

\end{document}